\documentclass[aps,nofootinbib,prd,superscriptaddress, tightenlines,
notitlepage, twocolumn, floatfix]{revtex4-1}
\usepackage{tabularx}
\usepackage{graphicx}
\usepackage{epsfig}
\usepackage{amssymb,bm,tensor}
\usepackage{textcomp}
\usepackage{amsmath}
\usepackage[varg]{txfonts}
\usepackage{enumerate}
\usepackage{mathtools}
\usepackage[normalem]{ulem}
\usepackage{bm} 
\usepackage[OT1]{fontenc}
\usepackage[colorlinks]{hyperref}
\usepackage{multirow}
\usepackage{makecell}
\usepackage{soul}
\usepackage{booktabs}
\usepackage{nameref}
\usepackage{threeparttable}

\usepackage[capitalize]{cleveref}
\newcommand{\dd}{{\rm d}}

\usepackage[dvipsnames]{xcolor}
\hypersetup{colorlinks=true,
            citecolor=NavyBlue,
            linkcolor=NavyBlue,
            urlcolor=NavyBlue}

\begin{document}
\title{Convective stability analysis of massive neutron stars formed in binary mergers}

\date{\today}

\author{Yong Gao}
\email{yong.gao@aei.mpg.de}
\affiliation{Max Planck Institute for Gravitational Physics (Albert Einstein Institute), 14476 Potsdam, Germany}

\author{Kota Hayashi}
\affiliation{Max Planck Institute for Gravitational Physics (Albert Einstein Institute), 14476 Potsdam, Germany}

\author{Kenta Kiuchi}
\affiliation{Max Planck Institute for Gravitational Physics (Albert Einstein Institute), 14476 Potsdam, Germany}
\affiliation{Center of Gravitational Physics and Quantum Information, Yukawa Institute for Theoretical Physics, Kyoto University, Kyoto, 606-8502, Japan} 

\author{Alan Tsz-Lok Lam}
\affiliation{Max Planck Institute for Gravitational Physics (Albert Einstein Institute), 14476 Potsdam, Germany}

\author{Hao-Jui Kuan}
\affiliation{Max Planck Institute for Gravitational Physics (Albert Einstein Institute), 14476 Potsdam, Germany}

\author{Masaru Shibata}
\affiliation{Max Planck Institute for Gravitational Physics (Albert Einstein Institute), 14476 Potsdam, Germany}
\affiliation{Center of Gravitational Physics and Quantum Information, Yukawa Institute for Theoretical Physics, Kyoto University, Kyoto, 606-8502, Japan}

\begin{abstract}

We perform fully general-relativistic hydrodynamics simulations of binary neutron star mergers over $100\,\rm ms$ post-merger to investigate the dynamics of remnant massive neutron stars (NSs). Our focus is mainly on the analysis of convective stability and mode characteristics of the massive NSs. We derive stability criteria for hot, differentially rotating relativistic stars that account for both buoyant and rotational restoring forces, and apply them for the first time to the post-merger massive NSs. Our results show no evidence of large-scale convective instability, as both angle-averaged specific entropy and specific angular momentum increase outward within the massive NSs. Rotational effects significantly enhance stability for local regions that would be otherwise unstable by the Schwarzschild criterion. Additionally, our mode analysis of matter fields and gravitational waves reveals no excitation of observable inertial modes after the damping of quadrupolar $f$-modes in the massive NSs, contrasting with previous studies. As in many previous works, we observe the excitation of an $m=1$ one-armed mode. However, we also find that the growth of the $m=1$ mode amplitude after the merger may correlate strongly with the violation of linear momentum conservation, indicating that we cannot reject the possibility that the excitation of the one-armed mode has a numerical origin.

\end{abstract}
\maketitle

\section{Introduction}

The simultaneous detection of gravitational waves (GWs) and electromagnetic radiation from the binary neutron star (BNS) merger event GW170817/GRB170817A/AT2017gfo marked the dawn of a new era in multimessenger astrophysics~\cite{LIGOScientific:2017vwq,LIGOScientific:2017zic,LIGOScientific:2017ync}. This landmark event demonstrated that BNS mergers are at least partial sources of short gamma-ray bursts, kilonovae, and the production of heavy elements via the rapid neutron-capture ($r$-process)~\cite{LIGOScientific:2017zic,LIGOScientific:2017ync,Pian:2017gtc,Goldstein:2017mmi,Savchenko:2017ffs,Mooley:2018qfh,Metzger:2010sy,Li:1998bw,Wanajo:2014wha,Eichler:1989ve,Shibata:2019wef,Radice:2020ddv}. 
{BNS mergers provide a unique opportunity to investigate the composition and state of supranuclear matter inside neutron stars (NSs). For the GW170817 event, the constraints on the equation of state (EOS) are primarily obtained from tidal imprints in the GWs during the late inspiral phase~\cite{LIGOScientific:2017zic,LIGOScientific:2018cki,LIGOScientific:2018hze}, which disfavor extremely stiff EOSs.}

To understand what GWs tell us about the matter properties requires modelling the merger dynamics of BNSs, a task that spans a few techniques. A post-Newtonian expansion of the two-body problem is well-suited for describing the early orbital phase, while numerical relativity becomes essential for accurately capturing the dynamics during the near-merger phase. In this stage, numerical relativity is being used to calibrate and promote perturbative approaches (e.g., effective one-body formalism), enabling the incorporation of matter effects—such as tidal deformation~\cite{Hinderer:2007mb,Flanagan:2007ix,Hotokezaka:2013mm} and mode resonance~\cite{Hinderer:2016eia,Kuan:2024jnw}—in a more concise and tractable manner. Reliable theoretical predictions of post-merger GWs and matter dynamics can only be obtained through general-relativistic hydrodynamics simulations (see e.g.,~\cite{Baiotti:2016qnr,Faber:2012rw,Shibata:2019wef} for reviews). Besides, physical processes grounded in numerical relativity, such as kilonova light-curve modelling, $r$-process nucleosynthesis calculations, and photon radiation transfer simulations are crucial for understanding the electromagnetic counterparts of BNS mergers.

The fate of the merger remnant depends primarily on the total mass and the EOS of NSs. 
{Numerical simulations show that the remnant can either collapse directly into a black hole or form a massive NS~\cite{Shibata:1999wm,Baumgarte:1999cq,Shibata:2005ss,Hotokezaka:2013iia}.} Observing GWs in either scenario can set an upper bound on the threshold mass for prompt collapse~\cite{Rezzolla:2017aly,Ruiz:2017due,Shibata:2019ctb} and reveals the matter dynamics of the massive NSs~\cite{Bauswein:2012ya,Hotokezaka:2013iia}. Postmerger simulations of the latter case, spanning a few tens of milliseconds, show a spectrum dominated by a main peak at $\sim 2$--$4\,\rm kHz$, with secondary peaks on either side~\cite{Takami:2014zpa,Soultanis:2021oia}.

Detailed mode analysis indicates that the main peak arises from the $m=2$ $f$-mode, while the secondary peaks may be attributed to the coupling between the $f$-mode and quasi-radial $m=0$ mode, or spiral deformations when the massive NS mass is relatively below the collapse threshold~\cite{Stergioulas:2011gd,Bauswein:2015yca}. The correlation between the main $f$-mode and premerger global properties (such as mass, radius, and tidal deformability) can provide constraints on the EOS of NSs~\cite{Bernuzzi:2014kca,Takami:2014zpa} {(see also the review~\cite{Baiotti:2016qnr} and references therein,)} particularly regarding phase transitions inside NSs~\cite{Bauswein:2018bma,Weih:2019xvw,Most:2018eaw}. However, potential systematic uncertainties from gravitational theories~\cite{Lam:2024azd} and thermal effects~\cite{Fields:2023bhs,Raithel:2023zml,Raithel:2019gws,Blacker:2023afl} must also be considered, and need further studies.

For the massive NS case, a key aspect of the long-term evolution is the presence of (magneto)hydrodynamic instabilities that shape the dynamics of the system. For example, Kelvin-Helmholtz instability can lead to significant amplification of magnetic fields at the merger~\cite{Anderson:2008zp,Kiuchi:2014hja,Kiuchi:2015sga,Kiuchi:2017zzg}, while magneto-rotational instability may drive the formation of large-scale magnetic fields through the $\alpha\Omega$ dynamo in the outer envelope of the massive NSs~\cite{Kiuchi:2023obe} (see \cite{Kiuchi:2024lpx} for a recent review). 
The hydrodynamical $m=1$ one-armed instability has been observed in many simulations of BNS mergers in the post-merger phase~\cite{Radice:2016gym,East:2015vix,East:2016zvv,Paschalidis:2015mla,Lehner:2016wjg,Radice:2023zlw}. This instability appears to be seeded at merger and grows exponentially and saturate within $10$--$20\,\rm ms$ after merger. It can emit GWs efficiently at half the frequency of $m=2$ $f$-mode~\cite{Lehner:2016wjg,Radice:2023zlw}, and possibly drives the spiral wind, which efficiently transports angular momentum outwards, and the ejecta of this component could reach $\gtrsim 10^{-2}\,M_{\odot}$~\cite{Nedora:2019jhl}. However, it is worth noting that the physical origin of the $m=1$ one-armed instability is still not clear.

Hydrodynamical instabilities often induce large-scale asymmetric fluid motions, potentially leading to significant GW emissions and providing insights into the thermal and rotational properties of the massive NSs. Since the massive NSs formed from BNS mergers are hot and differentially rotating, it is interesting to investigate whether fluid instabilities observed in simulations or perturbative studies of differentially rotating NSs --- such as bar-mode instability~\cite{Shibata:2000jt}, low $T/W$ instability~\cite{Centrella:2001xp,Shibata:2002mr,Shibata:2003yj,Saijo:2002nmv,Passamonti:2020yvh,Xie:2020udh}, Chandrasekhar-Friedman-Schutz (CFS) instability~\cite{Chandrasekhar:1970pjp,Friedman:1975ApJ,Kruger:2019zuz,Andersson:1997xt}, and convective instability~\cite{Camelio:2019rsz,Camelio:2020mdi} --- could develop within the massive NSs. Among these instabilities, convective instability may play an important role. 
This type of instability is frequently observed in supernova explosion simulations~\cite{Janka:2012wk,Janka:2006fh,Burrows:2020qrp}, which not only facilitates the transfer of energy to power the explosion but also generates substantial GW emissions~\cite{Ott:2008wt}.

Recent studies~\cite{DePietri:2018tpx,DePietri:2019mti} conducted $\pi$-symmetric hydrodynamical simulations of BNS mergers lasting $10^2\,\rm ms$ post-merger and found that the massive NSs develop convective instability based on the Schwarzschild criterion. Their simulations further suggested that this instability triggers inertial modes, which then persisted for several tens of milliseconds at frequencies of several kilohertz. Refs.~\cite{Camelio:2019rsz,Camelio:2020mdi} identified convective instability in hot, differentially rotating NSs with merger-like thermal and rotational profile, with convection timescales on the order of tens of milliseconds. By contrast, the general-relativistic neutrino-radiation hydrodynamics simulations of BNS mergers in Ref.~\cite{Radice:2023zlw} found that the massive NS remains stably stratified according to the Ledoux criterion.

Refs.~\cite{DePietri:2018tpx,DePietri:2019mti,Camelio:2019rsz,Camelio:2020mdi,Radice:2023zlw} primarily focused on thermal and/or compositional buoyancy. However, the rotational restoring force may be important for the massive NSs since they exhibit rapid and differential rotation~\cite{Shibata:2005ss,Takami:2014zpa}. Neglecting its contribution to the criteria for convection will consequently lead to incomplete conclusions. 
The convective instability due to differential rotation was first developed by Reyleigh~\cite{rayleigh_scientific_papers} for an incompressible fluid in a cylindrical container, and was later extended by Solberg~\cite{solberg1936mouvement} and H{\o}iland~\cite{hoiland1941avhandliger} for non-barotropic stars. It is commonly called the Solberg-H{\o}iland criterion in astrophysical literature, and has been frequently applied to study the convective stability of black hole disk~\cite{Abramowicz:2011xu,Sekiguchi:2010ja,Fujibayashi:2020qda}. For relativistic rotating stars, \citet{Bardeen:1970vja} gave a heuristic argument, while Abramowicz~\cite{Abramowicz:2004tw} provided an intuitive derivation for homentropic stars. \citet{Seguin:1975} gave a more precise and comprehensive treatment for hot and differentially rotating relativistic stars, including heat flow and viscosity.

In this article, we aim to investigate the convective stability of post-merger massive NSs using three-dimensional, fully general-relativistic simulations that accounts for both rotational and thermal restoring forces, as well as the effects of general relativity. Additionally, the mode characteristics will be analyzed in both the matter field and GWs. 
{We emphasize that the goal of this work is not to perform a direct comparison with Refs.~\cite{DePietri:2018tpx,DePietri:2019mti,Radice:2023zlw}. Our simulations employ numerical setups—such as the choice of EOS, binary masses, resolution, and hydrodynamical schemes—that differ in several important respects from those used in Refs.~\cite{DePietri:2018tpx,DePietri:2019mti,Radice:2023zlw}. These differences motivate an independent and self-consistent analysis rather than a one-to-one comparison.}
The paper is organized as follows: In Sec.~\ref{sec:methods}, we outline the analysis methods and numerical models employed in this study. In Sec.~\ref{sec:convection}, we present the main hydrodynamic results, focusing on the thermal and rotational properties of the massive NSs in \cref{sec:thermal_rotation}, and a detailed convective stability analysis in \cref{sec:stability}. The characteristics of the modes are explored in Sec.~\ref{sec:mode}, including the $m=1$ instability in \cref{sec:one_arm}, the dispersion relation and characteristic frequencies in \cref{sec:dispersion}, and the inertial mode problem in \cref{sec:inertial}. Finally, we summarize our findings in Sec.~\ref{sec:summary}. Throughout this paper, unless otherwise specified, we adopt geometric units with $c = G = 1$ where $G$ and $c$ are the gravitational constant and speed of light, respectively.

\section{Methods and numerical models}
\label{sec:methods}

\subsection{Matter field and equation of state}
\label{sec:eos}

The matter field is described by perfect fluid, with the energy-momentum tensor given by
\begin{equation}
    T_{\mu\nu} = \left[\rho(1+\epsilon)+p\right] u_\mu u_\nu+p g_{\mu\nu}\,.
\end{equation}
Here $u^{\mu}$ is the four-velocity, $g$ is the determinant of the spacetime metric $g_{\mu\nu}$, and the pressure $p$ is a function of the rest-mass density $\rho$ and the specific internal energy $\epsilon$ through the EOS. In this paper, we employ the piecewise polytropic parametrization~\cite{Read:2008iy} of three barotropic EOSs, namely APR4~\cite{Akmal:1998cf}, SLy4~\cite{Douchin:2001sv}, and MPA1~\cite{Muther:1987xaa}. 
{The zero-temperature pressure $p_{\rm c}$ and specific internal energy $\epsilon_{\rm c}$ are parameterized by piecewise polytropes consisting of four segments: The adiabatic index of the crust is set to a fixed value, wihle three more indecies are used to approximate the core region (see \cite{Read:2009yp,Uryu:2009ye,Taniguchi:2010kj,Hotokezaka:2013mm,Hotokezaka:2015xka,Kiuchi:2017pte,Kiuchi:2019kzt} for more details).
In each segment, the EOS is given as}
\begin{equation}
     p_{\mathrm{c}}(\rho) =   K_{i} \rho^{\Gamma_{\mathrm{i}}}\,, \quad \epsilon_{\mathrm{c}}(\rho) = \frac{K_{i} \rho^{\Gamma_{\mathrm{i}}-1}}{\Gamma_{\mathrm{i}}-1}+\Delta\epsilon_{i}\,, \quad (i=0,1,2,3)
\end{equation}
where $\Delta\epsilon_{i}$ is calculated by imposing the continuity of the pressure and specific internal energy at the transitional density $\rho_{i}$.
The EOS is fixed once the transitional density $\rho_{i}$, the polytropic index $\Gamma_{i}$, and the polytropic constant of $i=0$, $K_0$, are chosen. {Our choices of these parameters of the selected EOSs follow that of Ref.~\cite{Read:2008iy}.} 
To account for the shock heating during and after merger, we add a $\Gamma$-law thermal component to the cold part. The resulting total pressure and specific internal energy of the hybrid EOS can be written as 
\begin{align}
    \label{eq:p}
    & p=p_{\mathrm{c}}+p_{\mathrm{th}}=K_{i} \rho^{\Gamma_{\mathrm{i}}}+\rho (\epsilon-\epsilon_{\mathrm{c}})\left(\Gamma_{\mathrm{th}}-1\right) \,,\\
    \label{eq:eps}
    & \epsilon=\epsilon_{\mathrm{c}}+\epsilon_{\mathrm{th}}=\frac{K_{i} \rho^{\Gamma_{\mathrm{i}}-1}}{\Gamma_{\mathrm{i}}-1}+\Delta \epsilon_{i}+\epsilon_{\mathrm{th}}\,,
\end{align}
where $p_{\mathrm{th}}$ and $\epsilon_{\rm th}$ are the thermal contributions. In general, $\Gamma_{\rm th}$ depends on density, temperature, and composition~\cite{Constantinou:2015mna,Keller:2022crb,Keller:2020qhx}. Since it has been suggested that a constant $\Gamma_{\rm th}$ approximation is suitable for investigating the fate of the merger remnant~\cite{Bauswein:2010dn}, we follow Ref.~\cite{Kiuchi:2017pte} and take $\Gamma_{\rm th}=1.8$ in this work. The dependence of the post-merger evolution process on $\Gamma_\mathrm{th}$ is investigated in a separate paper (\citet{Han:2025pho}).

In this hybrid EOS, the specific enthalpy is given by  
\begin{equation}
    h =\frac{e+p}{\rho}=1-\left(\Gamma_i-1\right) \Delta \epsilon_i+\Gamma_{\text {th }} \epsilon_{\text {th }}+\Gamma_i \epsilon_{\rm c}\,,
\end{equation}
where the energy density $e$ is expressed as $\rho(1+\epsilon)$. From the first law of thermodynamics, 
\begin{equation}
    \label{eq:first}
    \dd e=\rho T \dd S+h \dd \rho\,, 
\end{equation}
we obtain the temperature $T$ and the specific entropy $S$ as
\begin{align}\label{eq:temperature}
    T= & \frac{m_{\rm n}}{k_{\rm B}}(\Gamma_{\rm th}-1)\epsilon_{\rm th}\,,\\
    S = & \frac{k_{\rm B}}{m_{\rm n}}\log \left[\frac{\left(\epsilon_{\rm{th}}\right)^{1 /\left(\Gamma_{\rm{th}}-1\right)}}{c_0 \rho}\right]\,,
\end{align}
respectively. Here $m_{\rm n}$ is the nucleon mass, $k_{\rm B}$ is the Boltzmann constant, and $c_0$ is an arbitrary constant with dimensions of the inverse of density. The adiabatic index for adiabatic perturbation is~\cite{Masaru:2016bk,Camelio:2019rsz,DePietri:2019mti}
\begin{equation}
    \label{eq:adiabatic_index}
    \Gamma_1  = \frac{\rho}{p}\left(\frac{\dd p}{\dd \rho}\right)_s = \frac{\rho}{p}\left[\left.\frac{\partial P}{\partial \rho}\right|_{\varepsilon}+\left.\frac{P}{\rho^2} \frac{\partial P}{\partial \varepsilon}\right|_\rho\right]  =\Gamma_{\text {th }}+\left(\Gamma_i-\Gamma_{\text {th }}\right) \frac{p_{\rm c}}{p}\,,
\end{equation}
which will be frequently used to analyze the criteria of convective stability. 

Armed with the two-parameter EOS $p = p(\rho, \epsilon) $, and the normalization of the four-velocity \( u^{\mu}u_{\mu} = -1 \), we still need to evolve five variables, which can be taken as the rest-mass density \( \rho \), the specific internal energy \( \epsilon \), and the three spatial components of the four-velocity \( u_{i} \). The state of the matter field is determined by the conservation equation of energy-momentum,
\begin{equation}\label{eq:T_conserve}
    \nabla_\mu T^{\mu}_{~\nu} = 0\,,
\end{equation}
and the conservation equation of the rest mass,
\begin{equation}\label{eq:restmass}
    \nabla_{\mu}(\rho u^{\mu}) = \partial_t\left(\rho \sqrt{-g} u^t\right) + \partial_i\left(\rho \sqrt{-g} u^i\right) = 0\,.
\end{equation}

\subsection{Numerical setup for the simulations}
\label{sec:model}

We employ the latest version of our numerical relativity code \texttt{SACRA-MPI}~\cite{Yamamoto:2008js,Kiuchi:2017pte,Kiuchi:2019kzt}. Specifically, we implement a moving puncture version of the Baumgarte-Shapiro-Shibata-Nakamura formalism with the fourth-order finite difference both in time and space to solve Einstein's equation~\cite{Shibata:1995we,Baumgarte:1998te,Campanelli:2005dd,Baker:2005vv}.
The Z4c prescription is also incorporated to damp out the constraint violation~\cite{Hilditch:2012fp}.  We also employ the Harten-Lax-van Leer contact (HLLC) approximate Riemann solver, as implemented in Ref.~\cite{Kiuchi:2022ubj}, to handle hydrodynamical evolution.

To cover a wide dynamical range, \texttt{SACRA-MPI} employs a Cartesian nested grid with the 2:1 refinement and imposes the mirror symmetry with respect to the orbital $z=0$ plane. For the binary NS simulations presented in this work, the grid consists of ten refinement levels: Each NS is covered by four finer comoving domains with six coarser refinement domains covering both NSs and centering at the center of mass of the binary system. Each domain is covered by a uniform cell-centered Cartesian grid with $2N\times2N\times N$ for $(x,y,z)$. The grid resolution at the refinement level $\rm lv$ is $\Delta x^{(\mathrm{lv})}=L^{(\mathrm{lv})} /(2 N)$, where $L^{(\mathrm{lv})}$ is the size of the corresponding domain on a given level along the $x$ and $y$ directions. The cell-centered grid structure allows us to apply the reflux prescription, aiding in the conservation of rest mass, with numerical errors remaining below $\mathcal{O}(10^{-7}\,M_{\odot})$ for long-term simulations lasting for $10^2\,\rm ms$ after the binary mergers in this work.

The initial data (ID) for the BNS is generated using the public spectral code FUKA~\cite{Grandclement:2009ju,Papenfort:2021hod}. In our IDs, the equal-mass BNS are in a quasi-equilibrium state with an initial separation of $44.3\,\rm km$, completing 4--5 orbits before merging. Using the orbital eccentricity reduction method implemented in the public version, FUKA produces IDs with residual eccentricities on the order of $10^{-3}-10^{-2}$. We presented the key parameters of the three equal-mass BNS systems in Table~\ref{tab:eos_parameter}. In the following, we mainly discuss the results of the canonical resolution with $N=86$,  while the results for different resolutions of the \texttt{APR4-135135} model will be presented in the \cref{sec:AppendixC}.

\begin{table}[]
    \centering
    \caption{The properties of the three equal-mass BNS systems studied in this work. Note that three different resolutions are set for the APR4 EOS. The columns present the gravitational mass ($M$) and the baryonic mass ($M_{0}$) of the individual stars, the angular velocity of the binary at the start of each simulation with the separation $44.3\,\rm km$, the grid resolution of the finest box, $\Delta x_{\rm finest}$, and the grid number $N$ for each refinement domain. The numerical models are named according to the EOS and the gravitational mass of the two NSs. For example, \texttt{APR4-135135} refers to the system with APR4 EOS, and the gravitational mass $m_{1}=m_{2}=1.35\, M_{\odot}$.}
    \begin{tabular}{ccccccc}
    \toprule
    EOS & $M\,[M_{\odot}]$ & $M_{0}\,[M_{\odot}]$ & $\Omega_{0}\,[\rm kHz]$ & $\Delta x_{\rm finest}\,[\rm m]$  & $N$\\
    \hline 
    APR4 & 1.35 & 1.50 & 1.82 & 225.0 & 62 \\
    APR4 & 1.35 & 1.50 & 1.82 & 162.2 & 86 \\
    APR4 & 1.35 & 1.50 & 1.82 & 136.8 & 102 \\
    SLy4 & 1.28 & 1.40 & 1.78 & 163.5 & 86 \\
    MPA1 & 1.35 & 1.49 & 1.82 &  180.3 & 86 \\
    \bottomrule
\end{tabular}
\label{tab:eos_parameter}
\end{table}

\subsection{Extraction of the gravitational waves}
\label{sec:gw}

GWs are extracted from the complex Weyl scalar $\Psi_4$ in the wave zone. At a given extraction radius $r_0$, $\Psi_4$ can be decomposed into $(l, m)$ modes using spin-weighted spherical harmonics $_{-2}{\rm Y}_{l m}(\theta, \phi)$, as expressed by 
\begin{equation}
    \label{eq:psi4}
    \Psi_4\left(t_{\mathrm{ret}}, r_0, \theta, \phi\right) = \sum \Psi_4^{lm}\left(t_{\mathrm{ret}}, r_0\right)\,_{-2}{\rm Y}_{l m}(\theta, \phi)\,.
\end{equation}
Here, the retarded time $t_{\mathrm{ret}}$ is defined as~\cite{Fiske:2005fx,Kiuchi:2017pte}:
\begin{equation}
    t_{\mathrm{ret}} \equiv t - \left[D + 2 m_0 \ln \left(\frac{D}{2 m_0} - 1\right)\right]\,,
\end{equation}
with the total mass $m_0 = m_1 + m_2$, and the proper radius of the extraction sphere $D \approx r_0 \left[1 + {m_0}/{2 r_0}\right]^2$. In this paper, we extract $\Psi_4$ at $r_0 = 480\,M_{\odot}$ and extrapolate it to infinity using Nakano's method~\cite{Nakano:2015rda}. The strain of GWs contributed by each mode is then calculated by integrating $\Psi_4^{lm,\infty}$ twice in time 
\begin{align} 
    h^{lm, \infty}\left(t_{\mathrm{ret}}\right) & =h_{+}^{lm, \infty}\left(t_{\mathrm{ret}}\right)-i h_{\times}^{lm, \infty}\left(t_{\mathrm{ret}}\right) \nonumber \\ 
    & =-\int^{t_{\mathrm{ret}}} \dd t^{\prime} \int^{t^{\prime}} \Psi_4^{lm, \infty}\left(t^{\prime \prime}\right) \dd t^{\prime \prime}\,.
\end{align}
We use the fixed frequency integration~\cite{Reisswig:2010di} to calculate $h^{lm, \infty}\left(t_{\mathrm{ret}}\right)$, namely 
\begin{equation}
    h^{lm, \infty}\left(t_{\mathrm{ret}}\right)=\int \dd f^{\prime} \frac{\tilde{\Psi}_4^{lm, \infty}\left(f^{\prime}\right)}{\left(2 \pi \max \left[f^{\prime}, f_{\mathrm{cut}}\right]\right)^2} \exp \left(2 \pi i f^{\prime} t_{\mathrm{ret}}\right)\,,
\end{equation}
where $\tilde{\Psi}_4^{lm, \infty}(f)$ is the Fourier transformation of $\Psi_4^{lm, \infty}(t)$, and the cut frequency $f_{\rm cut}$ of the low frequency band is taken as $0.8\, m \Omega_0 /2 \pi$ following~\cite{Kiuchi:2017pte,Kiuchi:2019kzt}. 
The effective amplitude of the Fourier spectrum is defined as~\cite{Hotokezaka:2011dh,Kiuchi:2017pte}
\begin{equation}
\tilde{h}_{\rm eff}^{lm}(f)\equiv f\sqrt{\frac{\left|\tilde{h}^{lm}_{+}(f)\right|^2+\left|\tilde{h}^{lm}_\times(f)\right|^2}{2}}\,,
\end{equation}
where $\tilde{h}^{lm}_{+}(f)$ and $\tilde{h}^{lm}_{\times}(f)$ are the Fourier transformation of the plus and cross polarization modes, respectively. Note that here we omit the angular dependence of the GW source. The angular dependence of the GW source can be easily recovered by the spin-weighted spherical harmonics (see \cref{eq:psi4}).
The luminosity of GWs of each mode can be calculated by~\cite{Ruiz:2007yx}
\begin{equation}
    L_{\rm GW}^{lm} = \frac{\dd E^{lm}}{\dd t} = \lim _{r \rightarrow \infty}\frac{r^2}{16 \pi} \left|\int^{\rm ret} \Psi_4^{lm, \infty}(t^{\prime}) \dd t^{\prime}\right|^2 \,.
\end{equation}

\subsection{Onset criterion of convection instability}
\label{sec:criterion}

Local convection instability arises when a displaced fluid element continues to move further away from its equilibrium position, driven by buoyancy that overpowers stabilizing effects such as the NS's self-gravity and pressure~\cite{Thorne:1966ApJ}. In rotating NSs, the restoring force induced by rotation should also be considered in addition to thermal or compositional buoyancy~\cite{Tassoul2000,Abramowicz:2004tw,Bardeen:1970vja,Fricke:1969,Goldreich:1967}. The massive NSs typically evolve to an approximately stationary state at $\sim 20$--$30\,\rm ms$ after the merger. {\it For simplicity, we analyze the merger remnants using the convection instability criterion for a stationary, differentially rotating, hot relativistic star.} We will adopt the methods formulated by~\cite{Fricke:1969,Seguin:1975,Bardeen:1970vja} to derive the necessary conditions for convective stability.

Assuming that the matter satisfies the circularity condition (i.e., no meridian fluid motions), the spacetime metric $g_{\alpha\beta}$ of an axisymmetric rotating star in equilibrium reads
\begin{equation}
    \dd s^2=-e^{2 \nu} \dd t^2+e^{2 \psi}(\dd \phi-\omega \dd t)^2+e^{2 \mu}\left(\dd \varpi^2+\dd z^2\right)\,,
\end{equation}
where the metric functions ($\nu$, $\psi$, $\omega$, and $\mu$ ) depend only on $\varpi$ and $z$ with $\varpi$ being the coordinate distance from rotational axis. $e^{\nu}$ is the lapse function, $\omega$ is the angular velocity of the zero-angular-momentum observer (ZAMO) at location $(\varpi,z)$, and $e^{\psi}$ is the circumferential radius of the spacetime. The four-velocity is decomposed as
\cite{Butterworth76}
\begin{equation}
    u^\alpha=u^t\left(t^\alpha+\Omega \phi^\alpha\right)\,,
\end{equation}
where $t^\alpha$ and $\phi^{\alpha}$ are the timelike and rotational Killing vectors, respectively, and $\Omega=\dd \phi /\dd t$ is the angular velocity of a fluid element seen by an observer at rest at infinity. The normalization of four-velocity $u_\alpha u^\alpha=-1$ determines $u^t$ as 
\begin{equation}
    u^t=\frac{e^{-\nu}}{\sqrt{1-v^2}}\,,
\end{equation}
where $v=e^{\psi-\nu}(\Omega-\omega)$ is the three-velocity of a fluid element measured by a ZAMO.

The projection of the energy-momentum conservation equation (\ref{eq:T_conserve}) along $u^{\mu}$ yields the relativistic Euler equations 
\begin{equation}
    \label{eq:euler}
    (g^{\mu\nu}+u^{\mu}u^{\nu}) \nabla_\nu p + (\epsilon+p) a^{\mu} =0\,,
\end{equation}
where $a^{\mu}= u^\nu \nabla_\nu u^\mu$ is the four-acceleration of the fluid element.
Combining the above equation with the first law of thermodynamics [\cref{eq:first}] gives an energy conservation law, 
\begin{equation}
    \label{eq:energy}
    \rho T u^{\mu}\partial_{\mu}S=0\,,
\end{equation}
which suggests that the specific entropy is conserved along the worldline of a fluid element. For later use, we present some conserved quantities along the fluid lines. Namely, the specific energy ($\mathcal{E} = -hu_{t}$) and the specific angular momentum per unit rest mass ($j = hu_{\phi}$) are conserved along fluid lines due to the presence of timelike and rotational Killing vectors, respectively. Their ratio,
\begin{equation}
\ell = \frac{j}{\mathcal{E}} = -\frac{u_{\phi}}{u_{t}}\,,
\end{equation}
is clearly conserved as well. This quantity represents the angular momentum per unit inertial mass and is independent of the thermodynamic properties of the fluid, and is crucial in determining the convective stability of differentially rotating relativistic stars~\cite{Seguin:1975,Kozlowski:1978}.

Although perturbations for relativistic stars generically involve both fluid variables and spacetime metric functions, the perturbation in the gravitational field can be ignored since convective stability is a local measure, i.e., perturbations are confined to an arbitrarily small region of the star or, in the case of axisymmetric perturbations, to an arbitrarily thin ring~\cite{Thorne:1966ApJ,Seguin:1975,Friedman:2013xza}. Therefore, we only need to consider the perturbation of fluid variables to derive the criteria. Specifically, we axisymmetrically perturb the fluid variables $\rho$, $\epsilon$, $u^{\varpi}$, $u^{z}$, and $u^{\phi}$. If $X$ represents one of theses variables, we can write 
\begin{equation}
    X(r, z, t)=X_{*}(r, z,t=0)+\delta X \exp \left[\sigma t-i\left(k_{\varpi} {\varpi}+k_z z\right)\right]\,,
\end{equation}
where $X_{*}$ is the value in the unperturbed configuration, $X$ is the perturbed value at the same position, and $\delta X$ is the Eulerian perturbation. In addition, $\sigma t$ represents the time-dependent perturbation, and the relevant components of the spatial wave vector (i.e., $k_{\varpi}$ and $k_z$) determine the total wavenumber as
\begin{equation}
    k^{2} = g^{\alpha\beta}k_{\alpha\beta} = e^{-2 \mu}\left(k_{\varpi}^2+k_z^2\right)\,.
\end{equation}
The local feature of the perturbations justifies the {\it short-wavelength approximation}
$\left(k R\right)^{-1}\ll 1$, where $R$ is the length scale of the star. We focus on perturbations having time scales comparable to the rotational time scale (i.e., $\sigma \sim \Omega$), for which the perturbations in the four-velocity satisfies 
\begin{equation}
    \label{eq:timescale}
    \delta u^{\varpi} \sim \delta u^z \sim R \delta u^\phi \sim \Omega / k\,,
\end{equation} 
and
\begin{equation}
    \label{eq:uphi}
    \delta u^{t} = - \frac{u_\phi}{u_t}\delta u^{\phi} = \ell \delta u^{\phi} \,.
\end{equation}

The variations in $\rho$ and $\epsilon$ are related to $\delta e$ and $\delta p$  by virtue of the EOS as   
\begin{align}
    \delta e & =(\partial e / \partial \rho)_\epsilon \delta \rho+(\partial e / \partial \epsilon)_\rho \delta \epsilon\,,\\
    \delta p & =(\partial p / \partial \rho)_\epsilon \delta \rho+(\partial p / \partial \epsilon)_\rho \delta \epsilon=0\,.
\end{align} 
Here $\delta p$ vanishes because the perturbation of the pressure can be quickly taken away on the timescale of sound-travel time $1/kc_{s}$. Since the Euler equations involve second-order derivatives, we also need the perturbations of the four acceleration~\cite{Seguin:1975},
\begin{align}
        & \delta a^t=\left(2 \nabla_{\varpi}u^t-\partial_{\varpi}u^t\right) \delta u^{\varpi}+\left(2 \nabla_{z}u^t-\partial_{z}u^t\right) \delta u^z+\sigma \ell u^t \delta u^\phi\,, \\
        & \delta a^\phi=\left(2 \nabla_{\varpi}u^\phi-\partial_{\varpi}u^\phi\right) \delta u^{\varpi}+\left(2 \nabla_{z}u^\phi-\partial_{z}u^\phi\right) \delta u^z+\sigma \ell u^t \delta u^\phi\,, \\
        & \delta a^\varpi=\sigma u^t \delta u^\varpi+e^{2(\psi+\nu-\mu)} u_t^{-3} \theta_\varpi \delta u^\phi\,, \\
        & \delta a^z=\sigma u^t \delta u^z+e^{2(\psi+\nu-\mu)} u_t^{-3} \theta_z \delta u^\phi\,,
\end{align}
where $\theta_i$ ($i=\varpi,z$) is defined by
\begin{widetext}
\begin{align}\label{eq:theta}
    {\theta}_{i}=\left(u^t u_t\right)^2\left[2(\Omega-\omega)(\partial_{i} \psi-\partial_{i} \nu)-\left(1+v^2\right) \partial_{i} \omega\right] = \left(u^t u_t\right)^2\left[\left(1-v^2\right) e^{-2 \psi} u_t^2 \partial_{i} \ell-\partial_{i} \Omega\right]\,.
\end{align}

By using the short wavelength approximation to eliminate higher order contributions, and by using the following equality 

\begin{equation}
    (e+p)\left\{g_{t \phi}\left(2 \nabla_{i} u^t-\partial_{i}u^t\right)+g_{\phi \phi}\left(2 \nabla_{i}u^\phi-\partial_{i}u^\phi{ }\right)\right\}+u_\phi \partial_{i}p=\rho\left(\partial_{i}\ell-u_\phi T \partial_{i}S\right)=u^t u_t^2(e+p) \partial_{i}\ell\,,
\end{equation}
we obtain the linear-perturbation equations corresponding to \cref{eq:restmass,eq:energy,eq:euler}. 
In particular, the rest mass conservation equation gives
\begin{equation}
    \label{eq:rest1}
    \sigma \delta\rho u^{t} /\rho -ik_{\varpi} \delta u^{\varpi} -ik_{z} \delta u^{z} =0\,;
\end{equation}
the energy conservation equation gives
\begin{equation}
    \label{eq:energy1}
    -\rho T (\partial_{\varpi} S \delta u^{\varpi} + \partial_{z}S\delta u^z)- \sigma u^t \left[(\partial e / \partial \rho)_\epsilon \delta \rho+(\partial e / \partial \epsilon)_\rho \delta \epsilon\right]+\sigma h u^t \delta \rho = 0 \,;
\end{equation}
while the spatial components of the relativistic Euler equations yield 
\begin{align}
    \label{eq:euler1}
    \sigma u^{t}e^{2\mu}(e+p)\delta u^{\varpi} + u_t^{-3}e^{2(\psi+\nu)}\theta_{\varpi}\delta u^{\phi}+a_{\varpi}\left[(\partial e / \partial \rho)_\epsilon \delta \rho+(\partial e / \partial \epsilon)_\rho \delta \epsilon\right] =&0 \,,\\
    \sigma u^{t}e^{2\mu}(e+p)\delta u^{z} + u_t^{-3}e^{2(\psi+\nu)}\theta_{z}\delta u^{\phi}+a_{z}\left[(\partial e / \partial \rho)_\epsilon \delta \rho+(\partial e / \partial \epsilon)_\rho \delta \epsilon\right] =&0\,,\\
    \label{eq:euler3}
    u^t u_t^2(e+p) \partial_{\varpi}\ell \delta u^{\varpi}+u^t u_t^2(e+p) \partial_{z}\ell \delta u^z-u^t u_t^{-1} e^{2(\psi+\nu)}\sigma u^t(e+p) \delta u^\phi=&0\,.
\end{align}
Eqs.~(\ref{eq:rest1}-\ref{eq:euler3}) comprise a set of five linear and homogeneous equations for the five quantities $\delta \rho$, $\delta \epsilon$, $\delta u^{\varpi}$, $\delta u^{z}$, and $\delta u^{\phi}$. The determinant of the coefficient matrix should vanish to have a non-trivial solution for the perturbed quantities. We then obtain an algebraic equation for $\sigma$,
\begin{equation}
    \label{eq:convect}
     \sigma^2+ e^{-4 \mu}\left(\frac{k_z}{k}\right)^2 \frac{1}{(u^t)^{2}} \Biggr[-\frac{1}{e+p}\left(\frac{\partial e}{\partial S}\right)_p\left(a_{\varpi}-\frac{k_\varpi}{k_z} a_z\right)\left(\partial_{\varpi}S-\frac{k_\varpi}{k_z} \partial_z S\right) +\left(\partial_\varpi\ell-\frac{k_\varpi}{k_z} \partial_z\ell\right)\left(\theta_\varpi-\frac{k_\varpi}{k_z} \theta_z\right) \Biggr]=0 \,.
\end{equation}
The roots of this equation determine the time evolution of the a perturbation with given values of $k_\varpi$ and $k_z$. To guarantee the convective stability of the system, a necessary condition is
\begin{equation}
    \label{eq:convect1}
      -\frac{1}{e+p}\left(\frac{\partial e}{\partial S}\right)_p\left(a_{\varpi}+\zeta a_z\right)\left(\partial_{\varpi}S+\zeta \partial_z S\right) +\left(\partial_\varpi\ell+\zeta \partial_z\ell\right)\left(\theta_\varpi+\zeta \theta_z\right) \geq 0 \,,
\end{equation}
where $\zeta = -{k_\varpi}/{k_z}$.
\end{widetext}

This condition implies that for the stability against the convection, a quadratic function of $\zeta$ must never be negative for any value of $\zeta$, which yields
\begin{align}
\label{eq:criterion}
    &\boldsymbol{\theta} \cdot \nabla \ell+(e+p)^{-2}(\partial e / \partial S)_p \nabla p \cdot \nabla S \geq 0\,,\\
    &(\partial e / \partial S)_p(\boldsymbol{\theta} \times \nabla p) \cdot(\nabla S \times \nabla \ell) \leq 0\,.
\end{align}
By using the thermodynamical relation 
\begin{equation}
    \nabla e  = (\partial e / \partial p)_S \nabla p+(\partial e / \partial S)_p \nabla S\,,
\end{equation}
we obtain two necessary conditions for the convective stability
\begin{align}
\label{eq:criterion1}
    & {\rm Criterion\  I:\quad}\boldsymbol{\theta} \cdot \nabla \ell+(e+p)^{-1} \nabla p \cdot \boldsymbol{A} \geq 0\,,\\
    \label{eq:criterion2}
    &{\rm Criterion\  II:\quad} (\boldsymbol{\theta} \times \nabla p) \cdot(\boldsymbol{A} \times \nabla \ell) \leq 0\,,
\end{align}
where $\boldsymbol{A}$ is the so-called Schwarzschild discriminant defined as
\begin{equation}
\label{eq:sch}
    \boldsymbol{A}=\frac{1}{e+p} \nabla e-\frac{1}{\Gamma_1 p} \nabla p\,.
\end{equation}
Here $\Gamma_1$ is the adiabatic index defined in \cref{eq:adiabatic_index}. One can recognize that Criterion I is the relativistic version of Solberg-H\o{}iland criterion~\cite{Tassoul2000}, which comes from the force balance between rotational and thermal effects, and has been frequently used in studying the convective stability of black hole disks~\cite{Abramowicz:2011xu,Sekiguchi:2010ja,Lattimer:1981cfw,Fujibayashi:2020qda}. This criterion can be easily understood in the following two limits. For non-rotating equilibria, the criterion reduces to the Schwarzschild criterion for stability~\cite{Thorne:1966ApJ}, i.e.,
\begin{equation}
    \nabla p \cdot\boldsymbol{A} \geq 0\,.
\end{equation}
When the criterion is met, a local fluid element will oscillate due to the buoyancy with the Brunt–V{\" a}is{\" a}l{\" a} (BV) frequency. To understand the impact of rotation on stability, one can consider isentropic fluid in equilibrium, for which the stable condition reduces to the relativistic Rayleigh-Solberg criterion~\cite{Abramowicz:2004tw,Bardeen:1970vja,Friedman:2013xza}
\begin{equation}
    \boldsymbol{\theta}\cdot \nabla \ell \geq 0 \,.
\end{equation}
Physical intuition can be built as follows: If a fluid element is perturbed outward axisymmetrically, it conserves its specific angular momentum $\ell$. When the ambient angular momentum decreases with radius, the fluid element will be rotating too fast for its new position, and the centrifugal force will push it further outward. On the other hand, the fluid element will oscillate at the local epicyclic frequency~\cite{Pringle:2007} if the stability condition holds.

It is necessary to perform coordinate transformation for relevant quantities before applying Criterion I and II that are derived from a quasi-isotropic cylindrical coordinate to the numerical results that are described on a Cartesian grid. In the 3+1 decomposition, the three-velocity $v^{i}$ and the Lorentz factor $W$ of a fluid element observed by an Eulerian observer can be written as 
\begin{align}
   v^i =&\frac{\gamma_\mu^i u^\mu}{-n_\mu u^\mu}=\frac{1}{\alpha}\left(\frac{u^i}{u^t}+\beta^i\right)\,, \\
    \quad W=& -n_{\mu}u^{\mu}=\alpha u^{t}= \frac{1}{\sqrt{1-v^2}}\,,
\end{align} 
where $\alpha$ is the lapse function, $\beta^i$ is the shift vector, $\gamma_{ij}$ is the spatial metric, and $n^{\mu}$ is the unit timelike vector normal to the hypersurface $\Sigma_t$. The angular velocity and the specific angular momentum are obtained as
\begin{align}\label{eq:Omega}
    \Omega \equiv & \frac{u^{\phi}}{u^t}=\alpha {v^{\phi}}-\beta^{\phi}\,,
\end{align}
and
\begin{align}
    \ell \equiv &  -\frac{u_{\phi}}{u_t} = -\frac{W v_{\phi}}{-\alpha W + \beta^{i}u_i}\,,
\end{align}
where $\beta^{\phi}$ is the shift vector in the $\phi$-direction. The quantities $v^{\phi}$, $\beta^{\phi}$, and $v_{\phi}$ can be expressed as  
\begin{align}
    v^{\phi}=&\frac{x v^y-y v^x}{{x^2+y^2}}\,,\\
    \beta^{\phi}=&\frac{x \beta^y-y \beta^x}{\sqrt{x^2+y^2}}\,,\\
    v_{\phi} =& \left(\frac{x v^y-y v^x}{x^2+y^2}\right)\left(y^2 \gamma_{x x}-2 x y \gamma_{x y}+x^2 \gamma_{y y}\right) \nonumber \\
    & +v^z\left(-y \gamma_{x z}+x \gamma_{y z}\right)\nonumber\\
    & + \left(\frac{x v^x+y v^y}{{x^2+y^2}}\right)\left({-y x} \gamma_{x x}-{y^2} \gamma_{x y}+\right. 
    \left.{x^2} \gamma_{y x} + {x y} \gamma_{y y}\right)\,.
\end{align}
The quantity $\theta_i$ [\cref{eq:theta}] can be rephrased as 
\begin{equation}
    \theta_{i}=\left(u^t u_t\right)^2\left[W^{-2} \gamma_{\phi\phi}^{-1} u_t^2 \partial_{i} \ell-\partial_{i} \Omega\right]\,,
\end{equation}
where the metric function $\gamma_{\phi\phi}$ can be written as 
\begin{equation}
    \gamma_{\phi\phi} = y^2 \gamma_{x x}-2 x y \gamma_{x y}+x^2 \gamma_{y y}\,,
\end{equation}
in terms of spatial metric in the Cartesian coordinates.

We have shown that the spatial distributions of $\Omega$ and $\ell$ are crucial for assessing the convective stability in axisymmetric stellar configurations.
While these two quantities are gauge-dependent, the dynamical gauge employed in simulations tend to preserve the essential characteristics of the minimal distortion gauge~\cite{Masaru:2016bk}. Moreover, Refs.~\cite{Kastaun:2014fna,Hanauske:2016gia} have investigated the influence of gauge conditions in detail and concluded that gauge artifacts have a minimal impact on rotational dynamics.
Thus, gauge effects are expected to be minor in the post-merger phase of BNSs, especially when the massive NS reaches a quasi-equilibrium state at approximately $20$--$30\,\rm ms$ after the merger. We will then disregard possible gauge-related artifacts in the following analysis.

\begin{figure*}
    \centering
    \includegraphics[width=16cm]{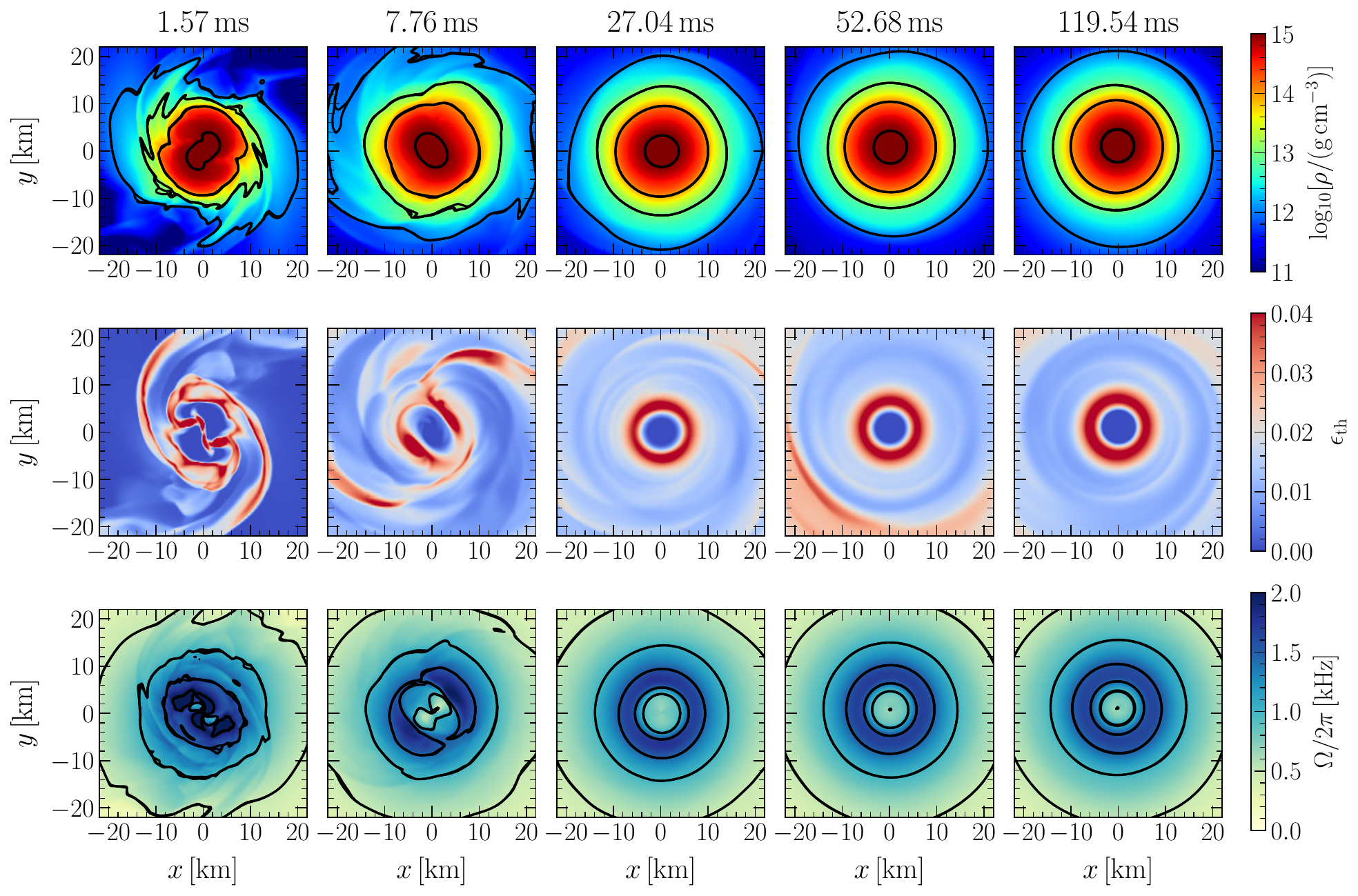}
    \caption{The distribution of rest-mass density (top), the thermal contribution to the specific internal energy (middle), and the fluid rotational frequency (bottom) in the equatorial $x$-$y$ plane is presented for the \texttt{APR4-135135} binary system at different post-merger times (see the top of the figure). The isocontours for the rest-mass density are plotted at ${\rm log}_{10}{[{\rho/(\rm g\,cm^{-3})}]}=\{12,13,14,15\}$, while those for the rotational frequency are shown at $\Omega/{(2\pi\,\rm kHz)}=\{0.5,1,1.5,2\}$. Note that the rotational frequency is not monotonic in the radial direction.
    }
    \label{fig:contour_reo}
\end{figure*}

\section{Convective stability of the merger remnants}
\label{sec:convection}

\subsection{Thermal and rotational properties}
\label{sec:thermal_rotation}

Since the criteria of convective instability are determined by the thermal and rotational properties of the massive NSs [cf.~Eqs.~(\ref{eq:criterion}-\ref{eq:criterion2})], we begin by examining the related quantities. \cref{fig:contour_reo} shows the 
spatial distribution of the rest-mass density $\rho$, the thermal component of the specific internal energy $\epsilon_{\rm th}$, and the rotational frequency $\Omega/2\pi$ on the equatorial $x$-$y$ plane at representative moments for the \texttt{APR4-135135} binary. As shown in the upper panels, a double-core structure is transiently formed in the first $\sim10\,\rm ms$ post-merger, making the massive NS an efficient emitter of GWs. Meanwhile, large amplitude quasi-radial oscillations induced during the merger lead to strong quasi-radial bouncing of the core. Since the kinetic energy is efficiently dissipated at the merger in the central region, the core rotates slower than the outer part, as illustrated in the second column of the bottom panel in Fig.~\ref{fig:contour_reo}. 

\begin{figure}
    \centering
    \includegraphics[width=\linewidth]{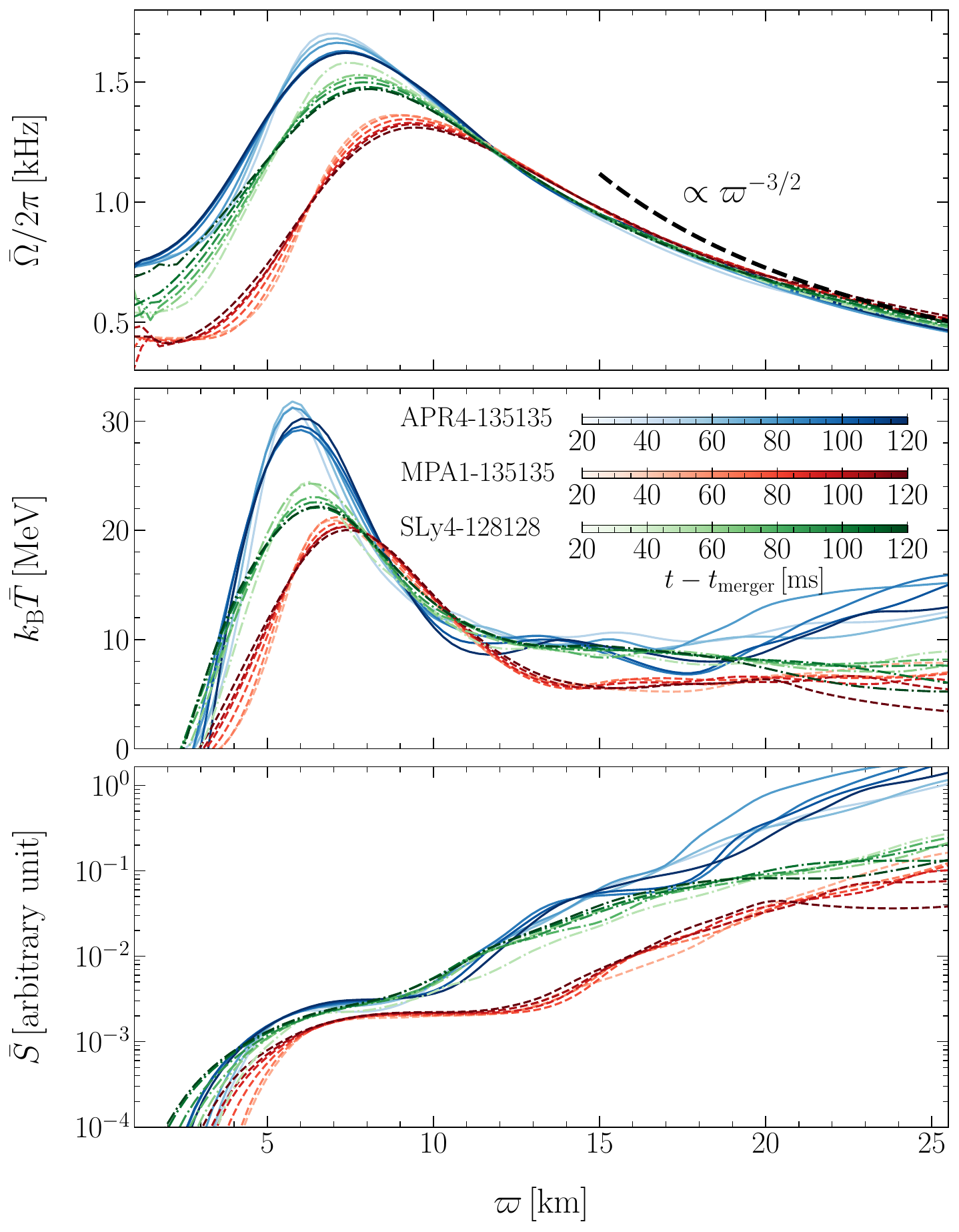}
    \caption{The angle-averaged rotational frequency $\bar{\Omega}/2\pi$ and thermal energy $k_{\rm B}\bar{T}$ for the \texttt{APR4-135135}, \texttt{MPA1-135135}, and \texttt{SLy4-128128} binaries at different after-merger times. The three subplots share a common set of labels displayed in the middle panel, with the ticks representing the post-merger time corresponding to the plotted curves.
}\label{fig:omega_T_S}
\end{figure}

In addition, the radial bounce induces shocks at the outer region of the massive NS during this stage, where the kinetic energy is efficiently converted to the thermal energy. This process produces a two ‘hot-spots’ structure in the specific internal energy around the core of the massive NS, which is closer to the center compared with the fast-rotating regions as depicted in the second column of the middle and bottom panels of \cref{fig:contour_reo} (see also \cref{fig:omega_T_S}). The shocks are also generated when the spiral arms hit the oscillating massive NS. 
On the other hand, the core of the massive NS is formed without experiencing the strong shock heating and can be approximately described by the cold EOS. From \cref{eq:temperature}, we obtain the thermal energy 
\begin{equation}
    k_{\rm B}T = 9.37(\Gamma_{\rm th}-1)\left(\frac{\varepsilon_{\mathrm{th}}}{0.01}\right)\,\mathrm{MeV}\,.
\end{equation}
The typical value of $\epsilon_{\rm th}=0.03$--$0.04$ at the shock region can then be translated to a thermal energy of $20$--$30\,\rm MeV$.

Since the spacetime rapidly approaches a stationary state and the deformation remains relatively small, it is sensible to consider azimuthal averages. For a quantity $X(t,r,\theta,\phi)$, the azimuthal-averaged value in the equatorial plane is evaluated as 
\begin{equation}
    \bar{X}(t,r)=\frac{1}{2 \pi} \int_0^{2 \pi} X(r, \theta=\frac{\pi}{2},\phi) \mathrm{~d} \phi\,.
\end{equation}
In Fig.~\ref{fig:omega_T_S}, we present the time evolution of the angle-averaged rotational frequency $\bar{\Omega}/2\pi$, thermal energy $k_{\rm B}\bar{T}$, and specific entropy $\bar{S}$ for the three binaries considered in this work. We start our analysis from $20$--$30\,\rm ms$ after the merger, and one can notice that the shown spatial distributions remain nearly unchanged until the end of our simulation since cooling and viscous effects are neglected in the present work. 

The rotational frequency becomes significantly low in the inner part ($\varpi \lesssim 3\, \mathrm{km}$) because the kinetic energy is efficiently dissipated at the merger in the central region. 
A sharp increase takes place at $\varpi \sim 5\text{--}10\,\mathrm{km}$, corresponding to the ring structure in Fig.~\ref{fig:contour_reo}. The peak frequency depends on the mass and radius of the NS in the binaries. For a given mass, as the case for \texttt{APR4-135135} and \texttt{MPA1-135135}, the softer EOS gives rise to a higher peak frequency as the merger remnant is more compact. The rotational frequency in the envelope decreases monotonically and scales as the Keplerian-law for the disk, i.e., $\propto \varpi^{-3/2}$ at $\varpi \gtrsim 15\,\mathrm{km}$. Note that the rotational frequency is quite similar in the envelope among the three cases. The rapid rotation in the envelope helps to balance the self-gravity of the massive NS as first pointed out in~\cite{Baumgarte:1999cq} for rotating NSs, and thoroughly studied in~\cite{Kastaun:2014fna,Kastaun:2016yaf} by merger simulations.

From the thermal energy distribution shown in the middle panel of Fig.~\ref{fig:omega_T_S}, one can observe that the central region ($\varpi \lesssim 3 \mathrm{~km}$) of the massive NSs is extremely cold. The peak of the thermal energy appears at larger radii, and compared to \texttt{MPA1-135135}, the maximal value of the thermal energy is higher for \texttt{APR4-135135}. The tendency indicates that the shock heating due to the quasi-radial bounce immediately after the merger is more efficient with increasing compactness of the NS. The envelope of the massive NS is repeatedly heated by the spiral arms, leading to a roughly uniform thermal energy in the envelope, on the order of $10\,\rm MeV$.

According to the above discussion, the massive NS approximately settles into an axisymmetric and quasi-equilibrium state at approximately $20$--$30\,\rm ms$ after the merger. 
At this stage, the criteria \eqref{eq:criterion1} and \eqref{eq:criterion2} derived in Sec.~\ref{sec:criterion} can be used to analyze the convective stability. Both Criterion I and II involve two pairs of quantities $\{p,\,\boldsymbol{A}\}$ and $\{\boldsymbol{\theta},\,\ell\}$ with the former determining the stability of spherical bodies against thermal buoyancy, and the latter determining the stability in isentropic rotating bodies.

\begin{figure}
    \centering
    \includegraphics[width=\linewidth]{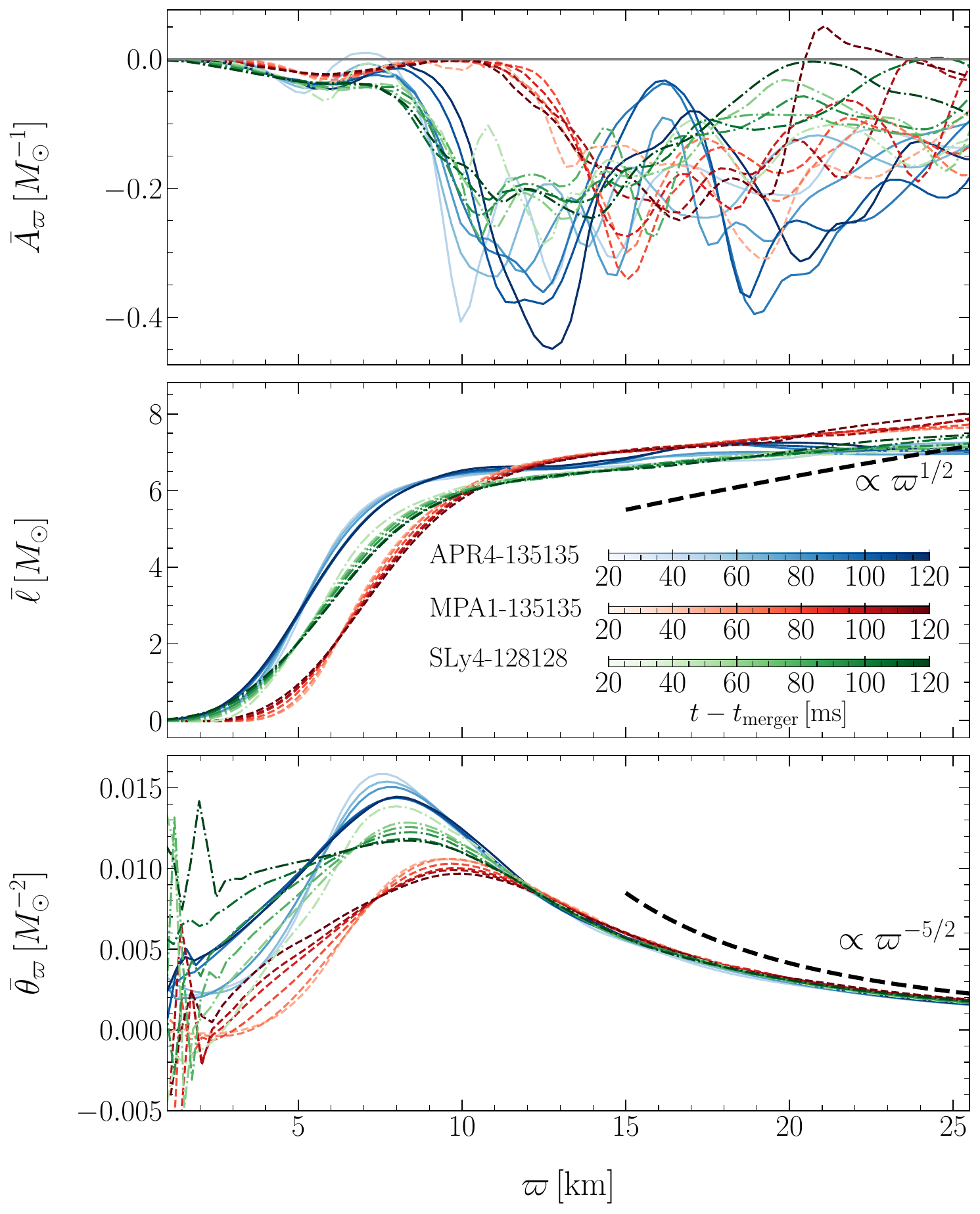}
    \caption{The time evolution of the angle-averaged radial component of the Schwarzschild criterion $\bar{A}_{\varpi}$ (top), the angle-averaged specific angular momentum $\ell$ (middle), and the angle-averaged radial component of the orientation vector $\theta_{\varpi}$ (bottom) on the equatorial plane for three different binary models. The horizontal gray line in the top panel indicates the place where $\bar{A}_{\varpi}=0$. The three subplots share a common set of labels displayed in the middle panel, with the ticks representing the post-merger times corresponding to the plotted curves.}
    \label{fig:theta_L}
\end{figure}

For simplicity, we examine the profiles of these quantities on the equatorial plane. As nothing interesting about the angle-averaged pressure $\bar{p}$, which decreases outward as expected for hydro-equilibrium, is observed, we will focus on $\boldsymbol{A},\,\boldsymbol{\theta}$, and $\ell$. Starting from the Schwarzschild criterion $A_{\varpi}$ in the top panel of \cref{fig:theta_L}, the generally negative value of such  ($A_{\varpi} < 0$) indicates stability against thermal buoyancy for all the simulated models, with exceptions observed in the model \texttt{APR4-135135} at $\varpi\sim 7\,\mathrm{km}$ during early post-merger and in the model \texttt{MPA1-135135} at $\varpi\sim 25\,\mathrm{km}$ in later post-merger stages. 

 We can look at this issue from another angle. Thermal convection normally happens when the specific entropy has a negative gradient. As shown \cref{fig:contour_reo,fig:omega_T_S}, the shock heating due to the quasi-radial bounce and later hydrodynamical interactions creates a hot ring around the core, thus resulting in a monotonic increase in specific entropy inside the massive NS. For the stellar exterior, the thermal energy decreases for larger radii. If the outer part is extremely cold, then the corresponding specific entropy will also have a negative gradient and trigger convection. However, this envelope region is heated up by the spiral arm continuously and has approximately uniform thermal energy, creating a positive entropy distribution. In conclusion, the shock heating due to quasi-radial bounce at merger time cannot produce a large enough specific entropy at the hot-ring region and the spiral arm heats the disk region of the massive NS up to high temperature. As a consequence, the entropy gradient is always positive (cf.~the lower panel of Fig.~\ref{fig:omega_T_S}), thereby preventing thermal convection.

We now analyze the effect of rotation from the angle-averaged specific angular momentum $\ell$ on the equatorial plane. As shown in the middle panel of Fig.~\ref{fig:theta_L}, the specific angular momentum $\ell$ increases outward, and the behavior can be explained by considering the rotation profile as follows.
Based on \cref{fig:omega_T_S}, the specific angular momentum should increase from the center to the location where the angular frequency reaches its maximum. Beyond that point, the angular frequency decreases following a Keplerian-like law ($\Omega \propto \varpi^{-3/2}$).
In Newtonian gravity, $\ell \rightarrow \Omega \varpi^2$, which yields the scaling of $\ell \propto \varpi^{1/2}$ as depicted in \cref{fig:theta_L}.

Refs.~\cite{Abramowicz:2004tw,Seguin:1975} show that for isentropic, stationary, and axially symmetric rotating fluids, the orientation vector $\boldsymbol{\theta}$ is always positively oriented.\footnote{Only in the vicinity of Kerr black holes can the strong curvature of spacetime cause the orientation vector to orient inward~\cite{Abramowicz:2004tw,Bardeen:1970vja}. However, in such cases, no equilibrium is possible for the fluid.} The radial component of the orientation vector $\theta_{\varpi}$ on the equatorial plane (bottom panel of \cref{fig:theta_L}) remains positive almost everywhere except for the central region of the massive NSs. In the Newtonian limit, the radial distribution of $\bar{\theta}_{\varpi} \rightarrow 2\Omega/\varpi$ closely resembles the rotational frequency shown in the upper panel of Fig.~\ref{fig:omega_T_S}. At large radii, it follows the scaling relation $\bar{\theta}_{\varpi} \propto \varpi^{-5/2}$. We suspect that the abrupt sign changes near the star’s center are caused by larger numerical errors in the differentiation of $\ell$ and $\Omega$. Additionally, the massive NS is not in complete equilibrium, primarily due to the presence of the $m=1$ instability. This issue will be further discussed in the next section.

\subsection{Stability analysis}
\label{sec:stability}

\begin{figure}
    \centering
    \includegraphics[width=\linewidth]{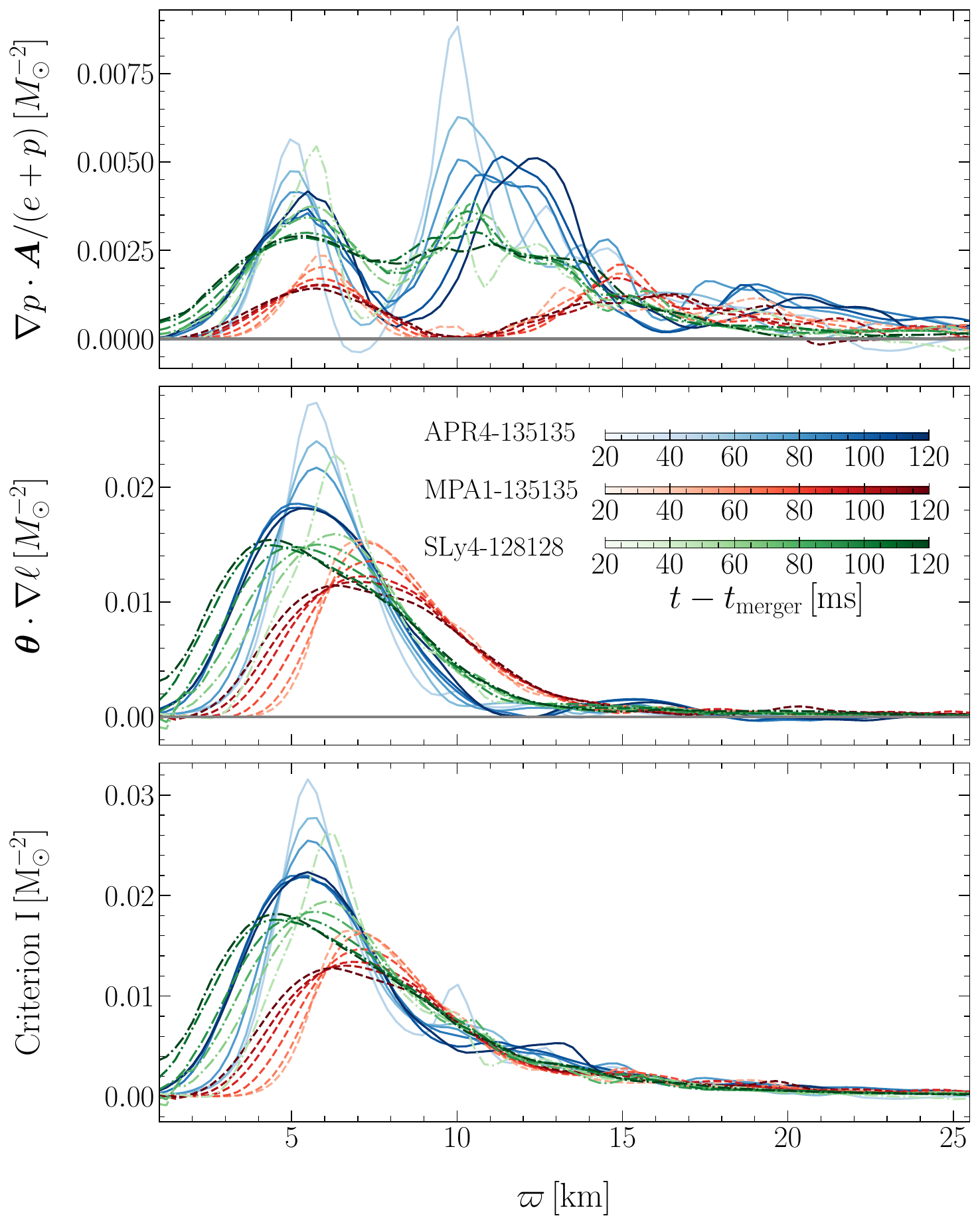}
    \caption{The time evolution of the angle-averaged Schwarzschild criterion, the Rayleigh-Solberg criterion, and the Criterion I on the equatorial plane for the three numerical models. The horizontal grey lines mark the place where the quantities are equal to zero. The three subplots share a common set of labels displayed in the middle panel, with the ticks representing the post-merger times corresponding to the plotted curves.}
    \label{fig:critI}
\end{figure}


Although we have separately examined the Schwarzschild and Rayleigh-Solberg criteria, the convective stability of massive NSs depends on Criterion I (relativistic Solberg-H{\o}iland criterion) as a whole and Criterion II. 
\cref{fig:critI} illustrates the time evolution of the angle-averaged Schwarzschild criterion, $\nabla p \cdot \boldsymbol{A}/(e+p)$, the relativistic Rayleigh-Solberg criterion, $\boldsymbol{\theta}\cdot \nabla \ell$, and Criterion I in the equatorial plane. The Schwarzschild criterion is nearly always satisfied, with weak violations being observed in the model \texttt{APR4-135135} at $\varpi\sim 7\,\mathrm{km}$ during the early post-merger phase and in the model \texttt{MPA1-135135} at $\varpi\sim 25\,\mathrm{km}$ during the later post-merger stages. Interestingly, this criterion exhibits an obvious double-peak structure across all three models considered in this study. By comparing with Fig.~\ref{fig:omega_T_S}, we observe that these peaks align with the inner and outer edges of the thermal energy peaks, where the specific entropy gradient is largest, thereby generating substantial buoyant forces. 

As shown in the middle panel of Fig.~\ref{fig:critI}, the Rayleigh-Solberg criterion is nearly always satisfied, except for weak violations at disk region with $\varpi \gtrsim 20\,\rm km$. The criterion peaks at $\varpi\sim 5$--$10\,\rm km$, where the specific angular momentum increases sharply (see Fig.~\ref{fig:theta_L}). In this region, the restoring centrifugal force per unit mass is largest. At small radii ($\varpi\lesssim 2\,\rm km$) and large radii ($\varpi\gtrsim 15\,\rm km$), the Rayleigh-Solberg criterion is approximately zero because the specific angular momentum is nearly constant at these regions, making the fluid element neutrally stable. It is clear that the rotational effects dominate over the thermal effects in the region with $\varpi \lesssim 10\,\rm km$ (see the scale of the vertical axis of Fig.~\ref{fig:critI}).

\begin{figure*}
    \centering
    \includegraphics[width=0.94\linewidth]{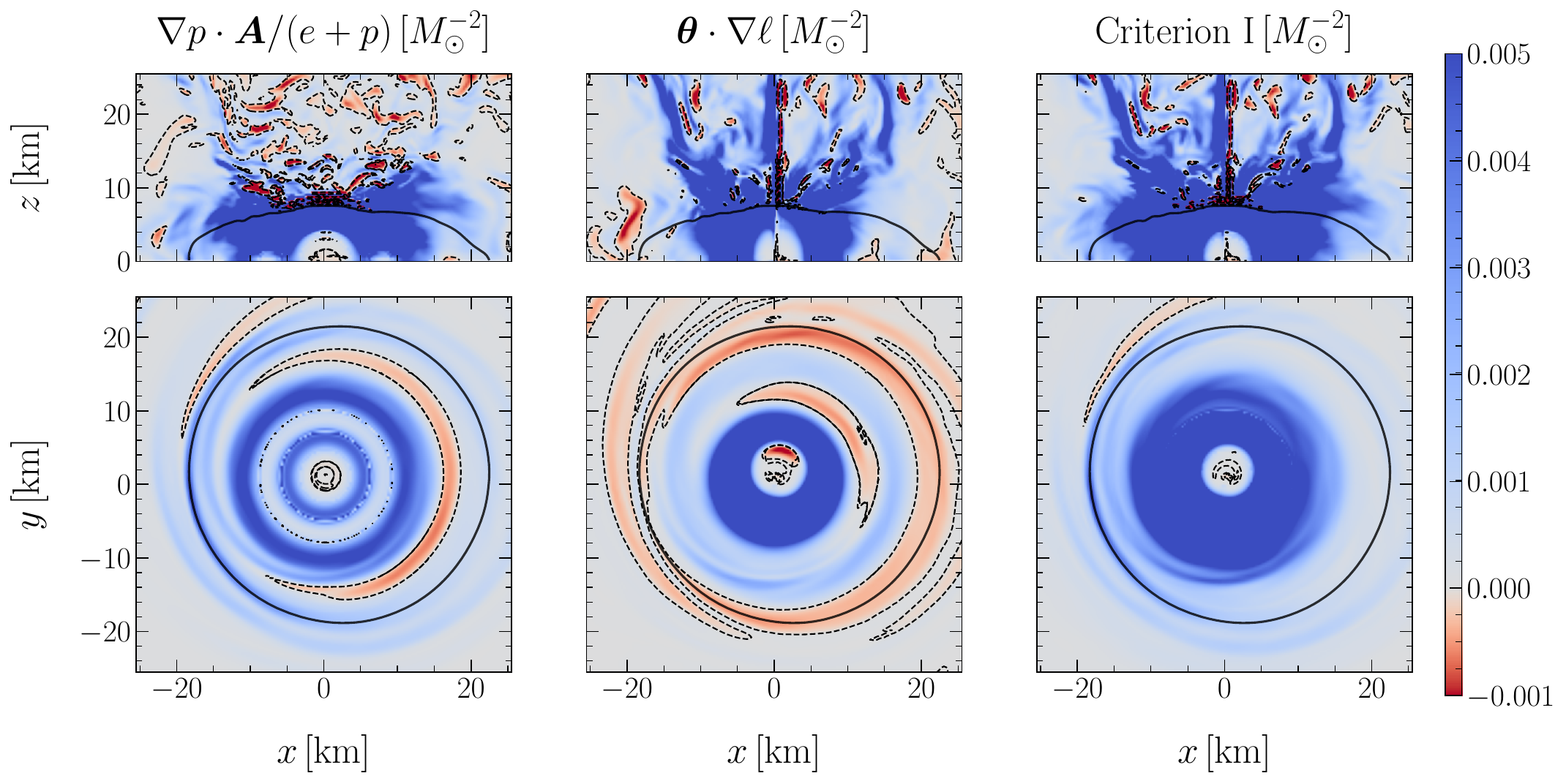}
    \caption{The snapshot of the Schwarzschild criterion value (first column), the Reyleigh-Solberg criterion value (second column), and the Criterion I value (third column) in the $x$-$z$ (first row) and $x$-$y$ (second row) planes for the model \texttt{APR4-135135} at post-merger time $t = 75.9\,\rm ms$. The black dashed curve marks the place where the criterion value equals to zero. The black solid enclosed curve represents the density contour at rest mass density $\rho =10^{12}\,\rm g\,cm^{-3}$.}
    \label{fig:critI_snapshot}
\end{figure*}

As shown in the bottom panel of Fig.~\ref{fig:critI}, the overall stability Criterion I is always satisfied because the rotational and thermal effects enhance the stability to each other for the massive NSs. One can notice that the region that violates the Schwarzschild criterion can be convectively stable because the restoring force due to rotation can suppress the instability caused by thermal buoyancy. Note that similar suppression of convection due to the centrifugal force has also been observed in the disk of black holes~\cite{Sekiguchi:2010ja,Fujibayashi:2020qda}. 

We also show the snapshot of the Criterion I in the $x$-$y$ and $x$-$z$ planes in Fig.~\ref{fig:critI_snapshot} for the model \texttt{APR4-135135} at post-merger time $t = 75.9\,\rm ms$. Both of the Schwarzschild criterion and the Reyleigh-Solberg criterion show instability regions in the $x$-$y$ planes, but the overall stability is guaranteed in most region. The central region of the massive NS has locations that violates the Criterion, which is mainly caused by numerical errors since the region is extremely cold and rotates slowly. In the $x$-$z$ plane, the polar region can violate the Schwarzschild criterion, but the rotational effect suppresses the overall instability heavily. The region that violates the Criterion I along the $z$ axis is likely to come from numerical errors. The region that violates the Criterion I in both the $x$-$z$ plane and $x$-$y$ plane locates at the place where the rest mass density $\rho \lesssim 10^{12}~{\rm g~cm^{-3}}$, and it does not have large impact on the central part if the instability really develops. Actually, we observe from the time evolution of the snapshot that the places that violate the criterion quickly change on the rotation timescale and the convection never develops globally. The animations of the {Criterion I} for the three numerical models can be found at \url{https://gravyong.github.io/convection/}. 
Our result is incompatible with Refs.~\cite{DePietri:2018tpx,DePietri:2019mti}, where the Schwarzschild criterion is first violated in the low density regions and quickly spreads out to the high density part of the massive NS.

\begin{figure}
    \centering
    \includegraphics[width=\linewidth]{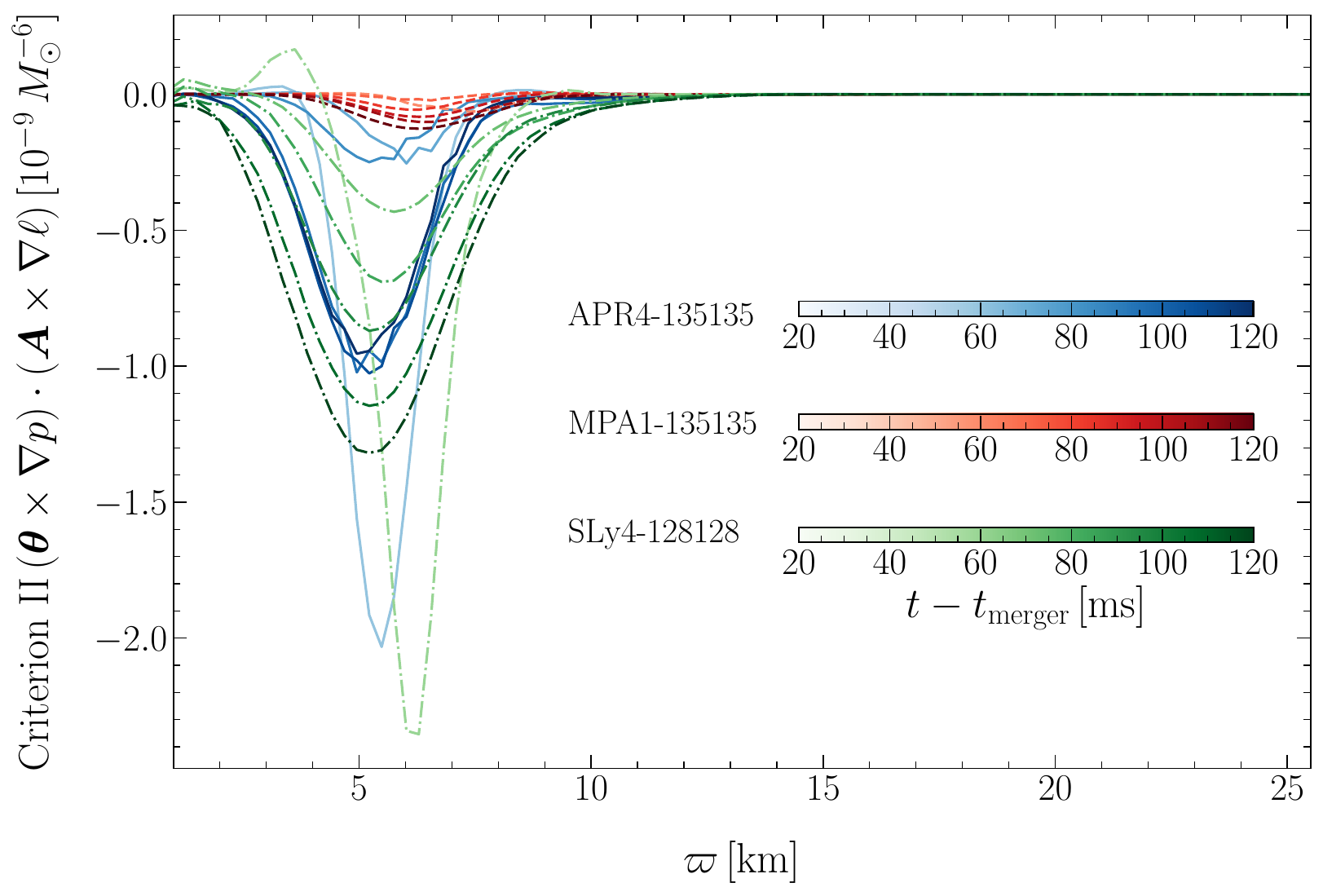}
    \caption{The angle-averaged Criterion II for the three numerical models. The ticks in the label represent the post-merger times corresponding to the plotted curves.}
    \label{fig:critII}
\end{figure}

Fig.~\ref{fig:critII} presents the angle-averaged Criterion II for three numerical models. This criterion is consistently satisfied, except for the {\texttt{SLy4-128128}} case, where it is violated approximately at $20\text{--}30\,\mathrm{ms}$ after the merger. The most negative region of Criterion II is located at $ \varpi \sim 5\,\mathrm{km}$ across all three models, caused by the steep gradient of the specific angular momentum. {Animations of Criterion II for the three numerical models are available at \url{https://gravyong.github.io/convection/}.}

Although Criteria I and II are only necessary conditions for convective stability (the violation of either criterion signifies a convective instability), they do serve as practical indicators of stability. 
{The key point is that these violations are weak and intermittent,
 and do not evolve into large-scale structures. 
 These observations strongly suggest that the system remains 
convectively stable.} A resolution study of the convective stability criteria is provided in \cref{sec:AppendixC}.

\section{Mode and GW analysis}
\label{sec:mode}

Hydrodynamical instabilities in the post-merger phase can have a large impact on both GWs and electromagnetic radiation.
Refs.~\cite{DePietri:2018tpx,DePietri:2019mti} report that the dominant quadrupolar $m=2$ $f$-mode diminishes at $30$--$50\,\rm ms$ after the merger, depending on the model. Afterward, the convective instability, as determined by the Schwarzschild criterion, excites global, discrete $m=2$ inertial modes that persist for tens of milliseconds. They proposed that these modes can be detected by third-generation GW detectors at frequencies of a few kilohertz, offering a unique opportunity to probe the rotational and thermal states of the merger remnant. Confirming this finding in independent simulations is critical to validating their existence.

\subsection{Putative $m=1$ excitation and concerns}
\label{sec:one_arm}

\begin{figure}
    \centering
    \includegraphics[width=\linewidth]{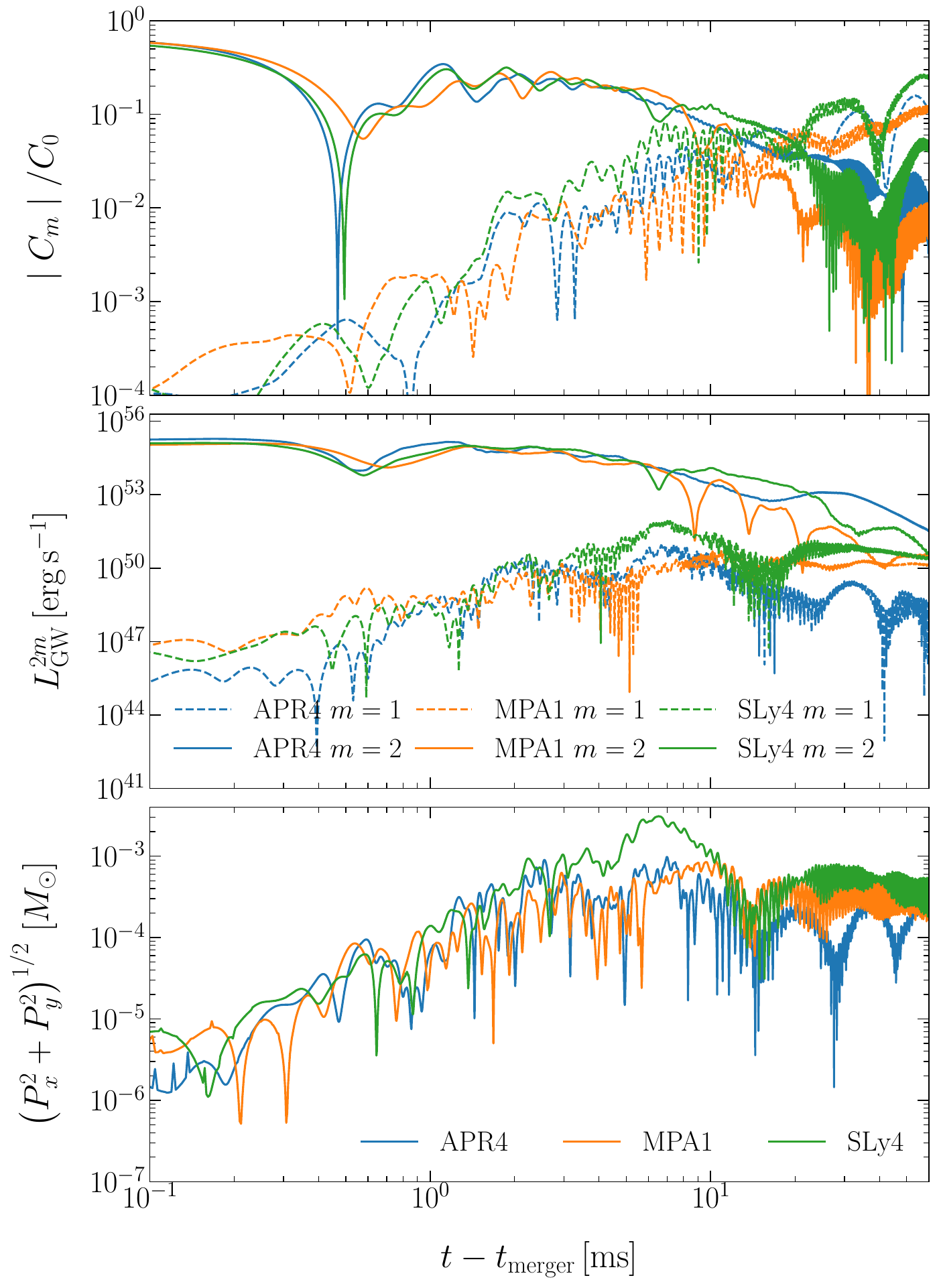}
    \caption{The post-merger time evolution of the normalized density projection, $|{C_{m}}|/{C_{0}}$ (top), the GW luminosity, $L_{\rm GW}^{2m}$ (middle), and the linear momentum violation, $\sqrt{P_x^2 + P_y^2}$ (bottom) for both $m = 1$ and $m = 2$ modes across the three numerical models. The top and middle panels share the same labels. Note that the post-merger time for the middle panel is the retarded value at $r_{0}= 480\, M_{\odot}$.
    }
    \label{fig:density_project}
\end{figure}

We quantitatively characterize the density configuration of the remnant massive NSs by computing the following modes,
\begin{equation}
    C_{m} = \int \rho W \sqrt{\gamma} e^{i m \phi}{\dd}^3 x\,,
\end{equation}
in the center-of-mass frame of the massive NS with the coordinate 
\begin{equation}
    x^i_{\rm cm} = \frac{1}{M_{0}}\int x^{i}\rho W \sqrt{\gamma} {\dd}^3 x\,.
\end{equation} 
In the top panel of Fig.~\ref{fig:density_project}, we depict the post-merger time evolution of $C_{1}$ (dashed) and $C_{2}$ (solid), normalized by $C_{0}$, for the three models. The $m=2$ mode dominates during the first $\sim10\,\rm ms$ but gradually decreases. In contrast, the $m=1$ mode seeded at the merger grows exponentially and saturates after $\gtrsim10$~ms has lapsed. It appears that the $m=1$ modes are not damped by hydrodynamical processes, and instead persist to the termination of simulations for all the considered models. This one-armed instability has been observed in many different BNS merger simulations~\cite{Radice:2016gym,East:2015vix,East:2016zvv,Paschalidis:2015mla,Lehner:2016wjg,Radice:2023zlw}. 

The excitation of modes will lead to the emission of GWs with the same azimuthal quantum number $m$. We display the luminosity of the $m=1$ and $m=2$ components of GWs in the middle panel of \cref{fig:density_project}. After the merger, the GW luminosity of the $m=2$ mode gradually decreases, while that of the $m=1$ mode grows exponentially and persists until the end of our simulations, although the luminosity of the $m=1$ mode is always subdominant. The behavior of both components closely resembles that of $C_{m}$. The correlation between the two is also seen in Refs.~\cite{Radice:2016gym,Lehner:2016wjg,Radice:2023zlw} for equal-mass BNS mergers.

Although the physical origin of the $m=1$ one-armed instability remains unclear, many studies~\cite{Radice:2016gym,East:2015vix,East:2016zvv,Paschalidis:2015mla,Lehner:2016wjg,Radice:2023zlw} have suggested that this instability is generic and may be observable with next-generation GW detectors. However, we emphasize two critical considerations that must be addressed before confidently interpreting the origin of this instability:

\begin{itemize} 
    \item It is essential to ensure that the numerical code used in simulations is free from bugs or artifacts that could artificially break the $\pi$-symmetry, leading to spurious results. 
    \item Any genuine physical instability dominated by the $m=1$ mode—such as the Papaloizou-Pringle instability~\cite{Papaloizou:1984MNRAS}—must inherently conserve linear momentum. Therefore, demonstrating linear momentum conservation is a necessary condition to establish the physical nature of the one-armed instability. 
\end{itemize}

To address the first point, we propose a symmetry-preserving test to demonstrate that the $\pi$-symmetry is perfectly maintained without explicitly imposing it under specific conditions. This test requires specifying a compiler option to control round-off errors. For Intel Fortran, the recommended option is \texttt{-fp-model strict}. The ID should exhibit $\pi$-symmetry to machine precision. We must also implement a symmetry-preserving technique described in \cref{sec:appendixA} in our numerical evolution code \texttt{SACRA-MPI}.

\begin{figure}
    \centering
    \includegraphics[width=\linewidth]{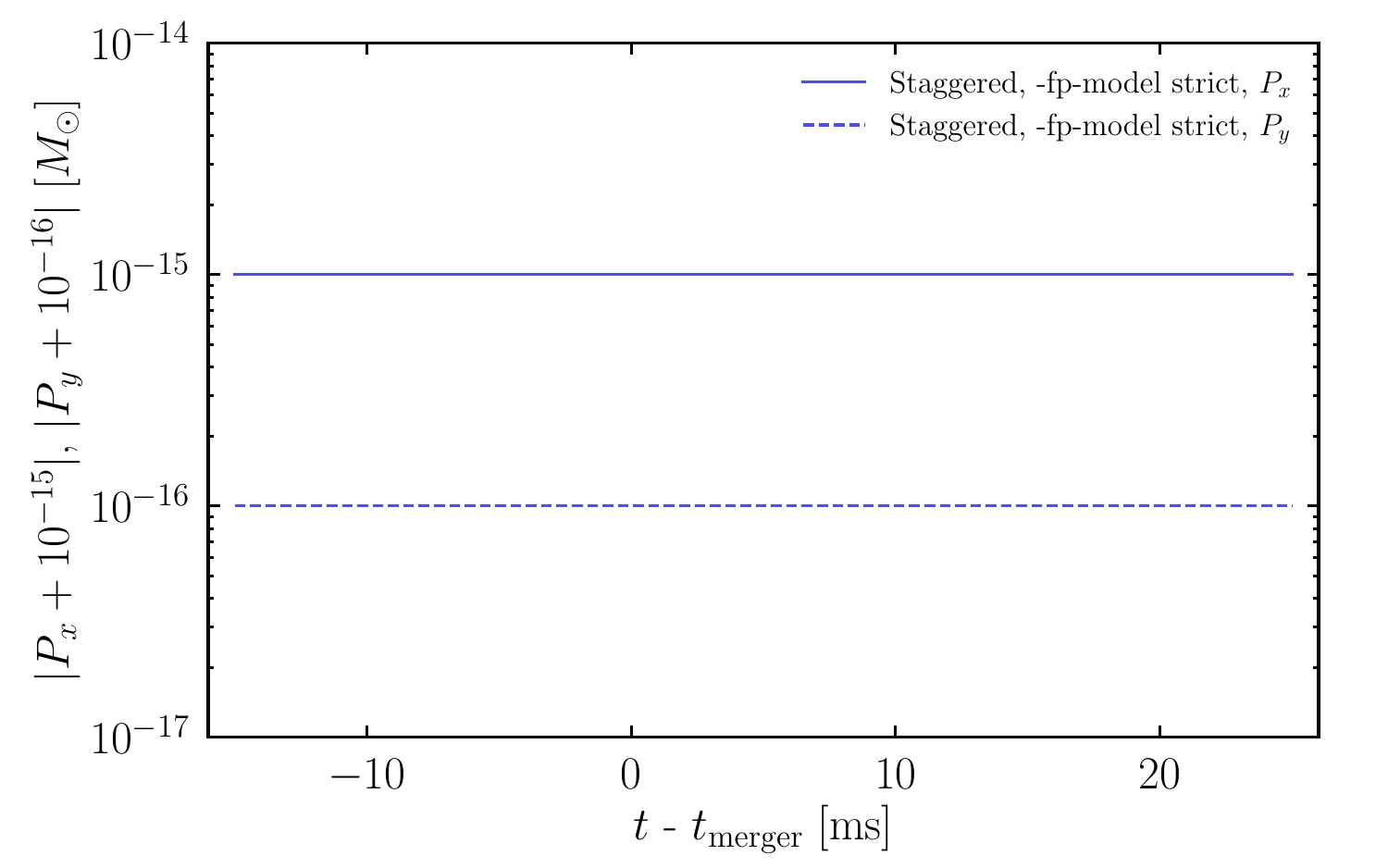}
    \caption{The linear momenta as functions of time are evaluated in the symmetry-preserving test for the \texttt{APR4-140140} model, with an initial orbital separation of $45\,\mathrm{km}$ and a grid resolution of $N=50$ for each refinement domain. To present the vertical axis on a logarithmic scale, a small offset is added: $10^{-15}$ for the solid line and $10^{-16}$ for the dashed line. The solid and dashed lines represent the $x$- and $y$-components of the linear momentum, respectively. }
    \label{fig:symmetry}
\end{figure}

\cref{fig:symmetry} plots the linear momentum in the $x$ and $y$ directions as functions of $t$ starting from the merger time. The linear momenta along the $x$ and $y$-axes are defined as
\begin{equation}
    P_{i} \equiv \int \sqrt{\gamma} \, \rho W h \, u_{i} \, {\dd}^3 x\,,~~~~i=x~\mathrm{or}~y.
\end{equation}
The $\pi$-symmetry is perfectly preserved from the inspiral to the post-merger phase. Therefore our code passes through the symmetry-preserving test. Because this test is necessary to explore the origin of the one-armed spiral instability, we strongly recommend the readers to perform this test.

As for the second point, the conservation of linear momentum in a simulation with a mesh refinement technique, such as the nested grid commonly used in the numerical relativity community, is generally non-trivial due to the mass going across the refinement boundaries after the merger (see \cref{sec:appendixB} for more detail). In the lower panel of Fig.~\ref{fig:density_project}, we present the linear momentum violation $\sqrt{P_x^2 + P_y^2}$ for the models considered in this work. Interestingly, the overall evolution of the violation of linear momentum conservation closely resembles that of the $m=1$ mode amplitude shown in the top panel as well as the $m=1$ GW luminosity presented in the middle panel. The numerical results obtained with different grid resolutions are presented in \cref{fig:resolution_projection} of \cref{sec:AppendixC}.  

It is observed that both the $m=1$ mode and the linear momentum violation grow exponentially after the merger, exhibiting a similar behavior regardless of the grid resolution. We evaluate growth timescale by fitting the data to an exponential function $A {\rm exp}\left(\sigma(t-t_{\rm merger
})\right)$. The results are shown in \cref{tab:m1} for the numerical models listed in \cref{tab:eos_parameter}, where $\sigma_{m}$, $\sigma_{\rm GW}$, and $\sigma_{\delta P}$ denote the growth rates of the $m=1$ mode, the luminosity of the GW in the $m=1$ channel, and the violation of the linear momentum. We set the post-merger time window starting from $t-t_{\rm merger}=0.1\,\rm ms$ to the position where the luminosity of the GW in the $m=1$ channel reaches its peak. 

Given two time series, $X$ and $Y$, the Pearson correlation coefficient, $C_p(X,Y)$, is defined as:
\begin{equation}
    {C}_P(X, Y)=\frac{\int_t\left(X-\langle X\rangle_t\right)\left(Y-\langle Y\rangle_t\right) \dd t}{\sqrt{\int_t\left(X-\langle X\rangle_t\right)^2 \dd t} \sqrt{\int_t\left(Y-\langle Y\rangle_t\right)^2 \dd t}}\,,
\end{equation}
where $\langle \rangle_t $ denotes the time average. To quantify the similarity, we compute the Pearson correlation coefficient $C_{p}\left({\rm log}(|C_{1}|/C_0),{\rm log}(\delta P)\right)$ between the $m=1$ mode and the linear momentum violation, as well as $C_{p}\left({\rm log}(L_{\rm GW}^{21}),{\rm log}(\delta P)\right)$ between the luminosity of the gravitational waves in the $m=1$ channel and the linear momentum violation in the selected time window. The results, listed in \cref{tab:m1}, show consistently high correlation values (larger than $0.80$ except for $C_{p}\left({\rm log}(|C_{1}|/C_0),{\rm log}(\delta P)\right)$ of \texttt{MPA1-135135} case), suggesting that the growth of the $m=1$ mode and the violation of linear momentum conservation might be highly correlated. Therefore, we cannot rule out the possibility that this instability has a numerical origin, at least for non-spinning symmetric BNS mergers reported in our study and in the literature~\cite{Radice:2016gym,Lehner:2016wjg,Radice:2023zlw}. 
Since the $m=1$ instability is not the central focus of this work, we will not delve deeper into it here, but the true origin of this instability warrants further investigation in future studies.

\begin{table*}[]
    \centering
    \setlength{\tabcolsep}{4.5pt}
    \caption{The growth rates along with their uncertainties, and Pearson correlation coefficients for different numerical models. The columns list the numerical model with grid resolution, the selected post-merger time window for fitting, the fitted values of $\sigma_{m}$, $\sigma_{\rm GW}$, and $\sigma_{\delta P}$ with their respective 1$\sigma$ uncertainties, as well as the Pearson correlation coefficients $C_{p}\left({\rm log}(|C_{1}|/C_0),{\rm log}(\delta P)\right)$, and $C_{p}\left({\rm log}(L_{\rm GW}^{21}),{\rm log}(\delta P)\right)$.}
    \begin{tabular}{cccccccc}
    \toprule
    Model & time window$\,[\rm ms]$& $\sigma_{m}\,[\rm ms^{-1}]$ & $\sigma_{\rm GW}\,[\rm ms^{-1}]$ & $\sigma_{\delta P}\,[\rm ms^{-1}]$ & $C_{p}\left({\rm log}\left(\frac{|C_{1}|}{C_0}\right),{\rm log}(\delta P)\right)$  & $C_{p}\left({\rm log}(L_{\rm GW}^{21}),{\rm log}(\delta P)\right)$\\
    \hline 
    \texttt{APR4-135135} (N=62) &$0.1\text{--}7.1$ & 1.43 $\pm$ 0.02 & 2.97 $\pm$ 0.03 & 1.61 $\pm$ 0.02 & 0.91 & 0.94 \\
    \texttt{APR4-135135} (N=86) & $0.1\text{--}7.4$& 1.59 $\pm$ 0.01 & 3.13 $\pm$ 0.03 & 1.15 $\pm$ 0.01 & 0.80 & 0.80 \\
    \texttt{APR4-135135} (N=102) & $0.1\text{--}3.7$& 1.57 $\pm$ 0.02 & 3.00 $\pm$ 0.03 & 1.47 $\pm$ 0.02 & 0.80 & 0.85 \\
    \texttt{SLy4-128128} (N=86) & $0.1\text{--}7.0$& 1.92 $\pm$ 0.01 & 3.53 $\pm$ 0.02 & 1.71 $\pm$ 0.01 & 0.95 & 0.93 \\
    \texttt{MPA1-135135} (N=86) &$0.1\text{--}10.9$ & 1.17 $\pm$ 0.01 & 1.67 $\pm $ 0.02 & 1.07 $\pm$ 0.01 &  0.74 & 0.81 \\
    \bottomrule
\end{tabular}
\label{tab:m1}
\end{table*}

\subsection{Characteristic frequencies}
\label{sec:dispersion}

The inertial modes can be classified as either polar-led gravito-inertial modes~\cite{Lockitch:2000aa,Unno1979} or axial-led inertial modes that reduce to the $r$-modes in the slow rotation limit~\cite{Andersson:1997xt}. The polar-led modes have a mixed inertia-gravity character and are restored by both Coriolis force and buoyancy inside rotating and stratified NSs~\cite{Passamonti:2008sb,Gaertig:2009rr}. 
They typically behave as $g$-modes in the slowly rotating limit, where buoyancy dominates the Coriolis force. For higher rotation rates, the effect of the Coriolis force is enhanced and eventually overpowers the buoyancy beyond a given stellar spin. In such a regime, this class of modes exhibits the properties of the inertial modes of barotropic models.

The mode classifications of merger-remnant massive NSs are still lacking (but see Ref.~\cite{Passamonti:2020yvh} for a Newtonian study). As a preliminary effort, we analyze the mode characteristics on the equatorial plane using dispersion relations. Considering that a fluid ring is slowly displaced from $\varpi_{1}$ to $\varpi_2 = \varpi_1 + \Delta \varpi$ in an axisymmetric manner while maintaining a fixed specific angular momentum $\ell$, the total radial acceleration experienced by the ring is
\begin{equation}
   \mathcal{A}_{\varpi} = -\gamma_{\varpi\varpi}^{-1/2} \left[\theta_{\varpi}\partial_{\varpi}\ell + (e+p)^{-1}\partial_{\varpi}p\, A_{\varpi}\right]\Delta{\varpi}\,.
\end{equation}
An acceleration directed away from the original position $(\mathcal{A}_{\varpi}/\Delta\varpi>0)$ indicates convective instability, while one directed back ($\mathcal{A}_{\varpi}/\Delta\varpi<0$) suggests stability [cf.~Criterion I \eqref{eq:criterion1}]. The corresponding angular frequency of the oscillation, 
\begin{equation}
    \omega^2 = -\mathcal{A}_{\varpi}/\Delta\varpi= N^{2} + \kappa^2\,,
\end{equation}
is determined by the relativistic BV frequency,
\begin{equation}
    N^{2} =\gamma_{\varpi\varpi}^{-1/2} (e+p)^{-1}\partial_{\varpi}p\, A_{\varpi}\,,
\end{equation}
and the relativistic epicyclic frequency,
\begin{equation}
    \kappa^2 =  \gamma_{\varpi\varpi}^{-1/2}\theta_{\varpi}\partial_{\varpi}\ell\,,
\end{equation}
in the local frame of the fluid ring.
The former characterizes the $g$-mode oscillations~\cite{Finn1986,Ipser1992,Yoshida:2012mu}, while the latter characterizes the inertial mode oscillations~\cite{Abramowicz:2011xu,Kato2001}. In Newtonian gravity, the above expression for $\kappa$ reduces to $\kappa^2=2 \Omega(2 \Omega+\varpi \dd \Omega / \dd \varpi)$, which becomes $\kappa=2\Omega$ for the uniform rotation and $\kappa=\Omega$ for the Keplerian rotation.

\begin{figure}
    \centering
    \includegraphics[width=\linewidth]{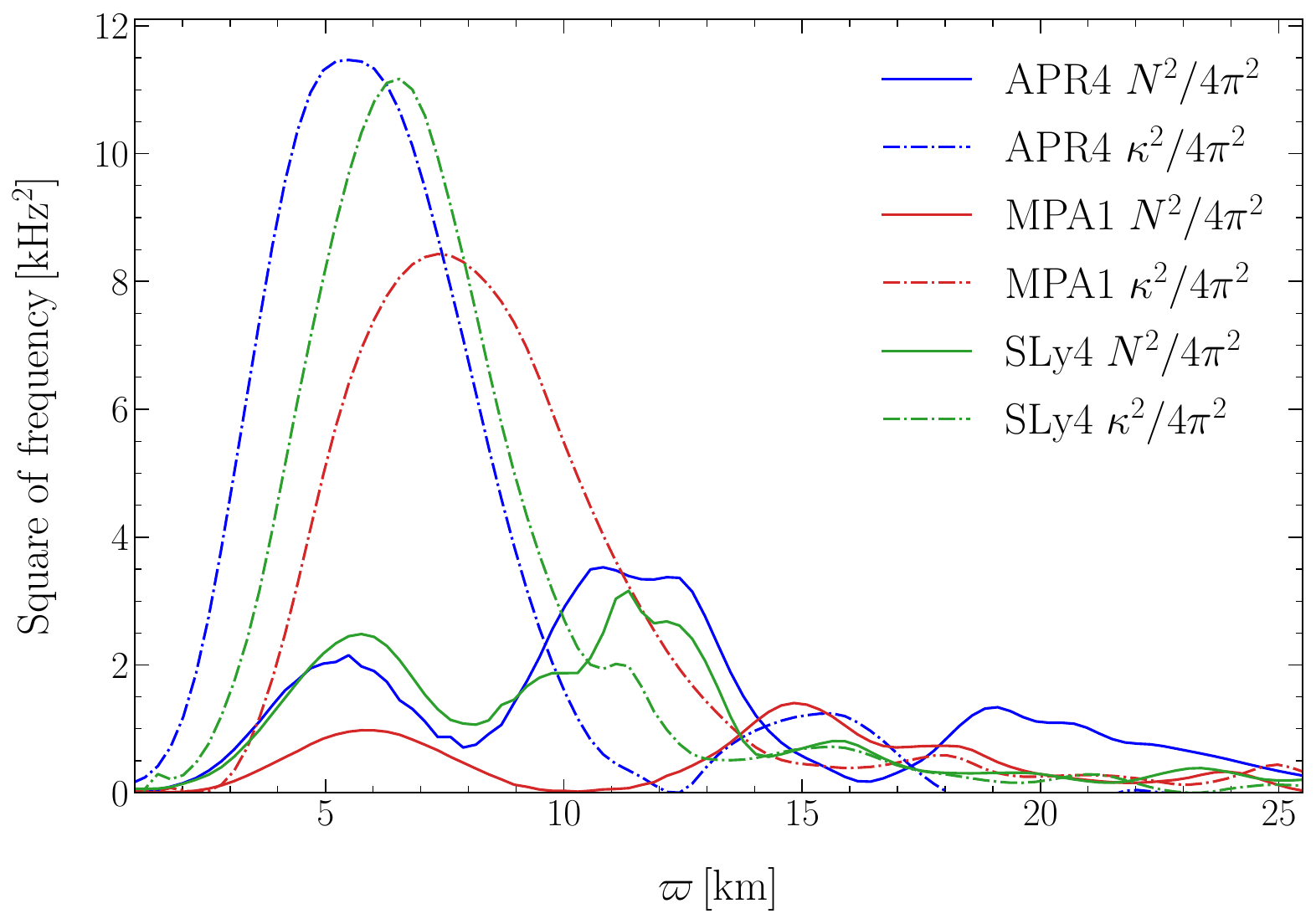}
    \caption{The angle-averaged BV frequency (solid) and the epicyclic frequency (dashed) on the equatorial plane for the three numerical models at post-merger time $t=80\,\rm ms$. Here we divide the quantities by $4\pi^2$ to represent the values for the square of frequency. }
    \label{fig:frequency}
\end{figure}

Fig.~\ref{fig:frequency} shows the angle-averaged profiles of $N$ and $\kappa$ on the equatorial plane for the three models at $t=80\,\rm ms$ after the merger. The square of the BV frequency exhibits two peaks around the maximum of the thermal energy shown in Fig.~\ref{fig:omega_T_S} where the entropy gradient is largest.  Another peak around $\varpi \simeq 18\,\rm km$ for the APR4 model can be attributed to the heating efficiency increases around this place as shown in Fig.~\ref{fig:omega_T_S}. The epicyclic frequency has one peak around the maximum of the rotational frequency. The radial dependence of the frequencies is similar to the top and middle panels of Fig.~\ref{fig:critI} because the Criterion I results from the force balance equations. The epicyclic frequency is much larger than the BV frequency in the region of $\varpi \lesssim 10\,\rm km$, by the factor of $\sim 6$--$8$. This indicates that the Coriolis force dominates over the thermal buoyant force. Around the second peak of the BV frequency, the rotational restoring force decreases sharply, and the BV frequency is larger than the epicyclic frequency; i.e., these two frequencies become comparable (except for APR4 model), and then, they approach zero at larger radii $\varpi \gtrsim 20\,\rm km$.

Overall, the oscillations of inertial modes, including polar-led gravito-inertial modes and axial-led inertial modes, in most regions of the massive NS are primarily influenced by rotational effects. As shown in Fig.~\ref{fig:frequency}, the characteristic frequency, $ \omega/2\pi $, of the inertial modes ranges from approximately $ 2.3 \text{--}3.4\,\mathrm{kHz}$ within the massive NS, depending on the EOS and total mass. This rough estimate aligns with previous perturbative studies for uniformly rotating NSs~\cite{Passamonti:2008sb,Passamonti:2020yvh,Gaertig:2009rr} and numerical simulations~\cite{DePietri:2018tpx,DePietri:2019mti}. The question, however, is whether the inertial mode can reach a substantial amplitude without the presence of convective instability. We will address this in the following section.

\subsection{No excitation of observable inertial mode}
\label{sec:inertial}

\begin{figure}
    \centering
    \includegraphics[width=\linewidth]{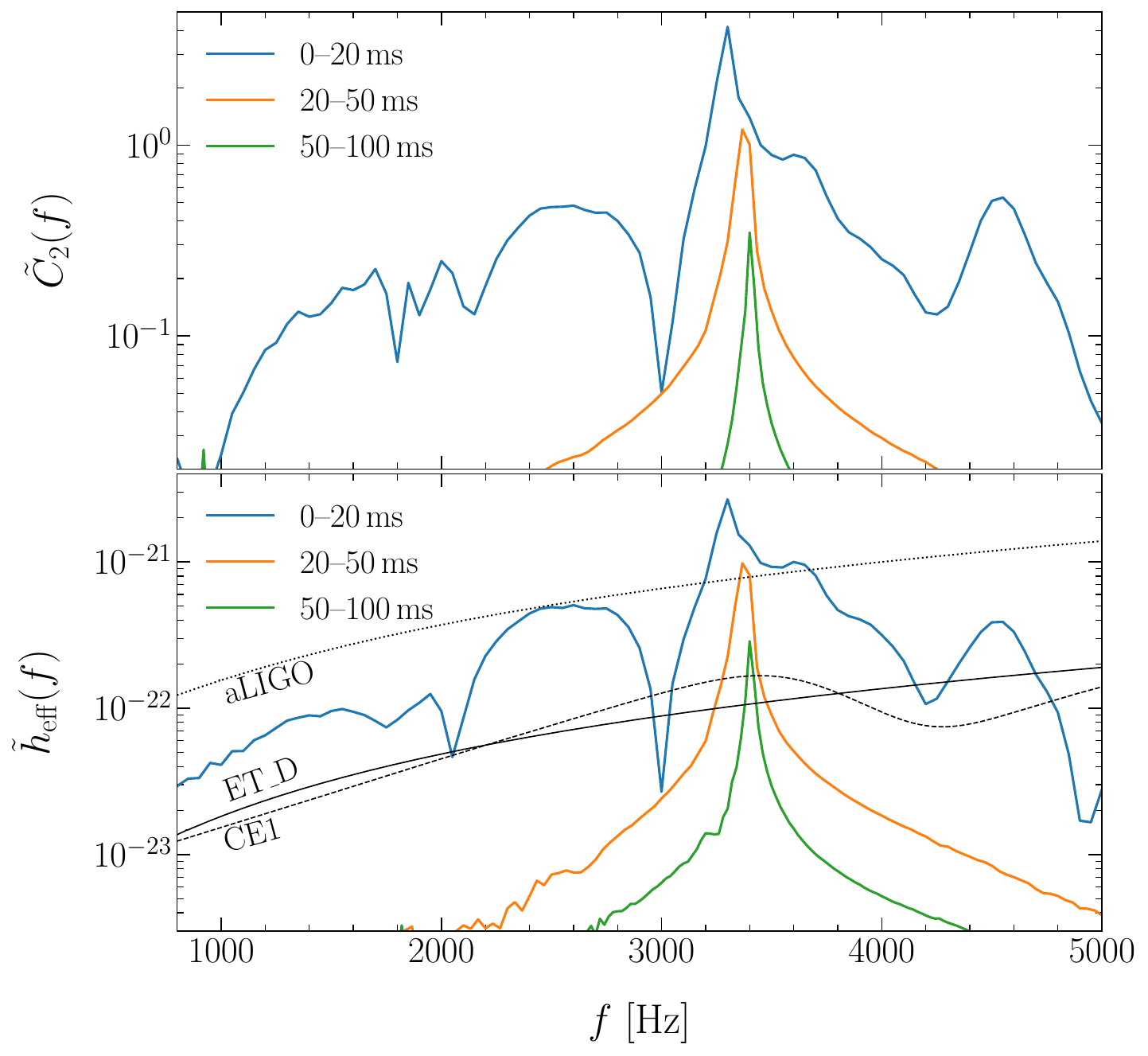}
    \caption{The amplitude spectrum density (ASD) of the mass projection for the $l=m=2$ mode $\tilde{C_2}(f)$ (top) and the effective amplitude of GWs $\tilde{h}_{\rm eff}(f)$ (bottom) for the \texttt{APR4-135135} model at different post-merger time window. The luminosity distance of the BNS is taken as $50\,\rm Mpc$.}
    \label{fig:waveform}
\end{figure}

\begin{figure*}
    \centering
    \includegraphics[width=\linewidth]{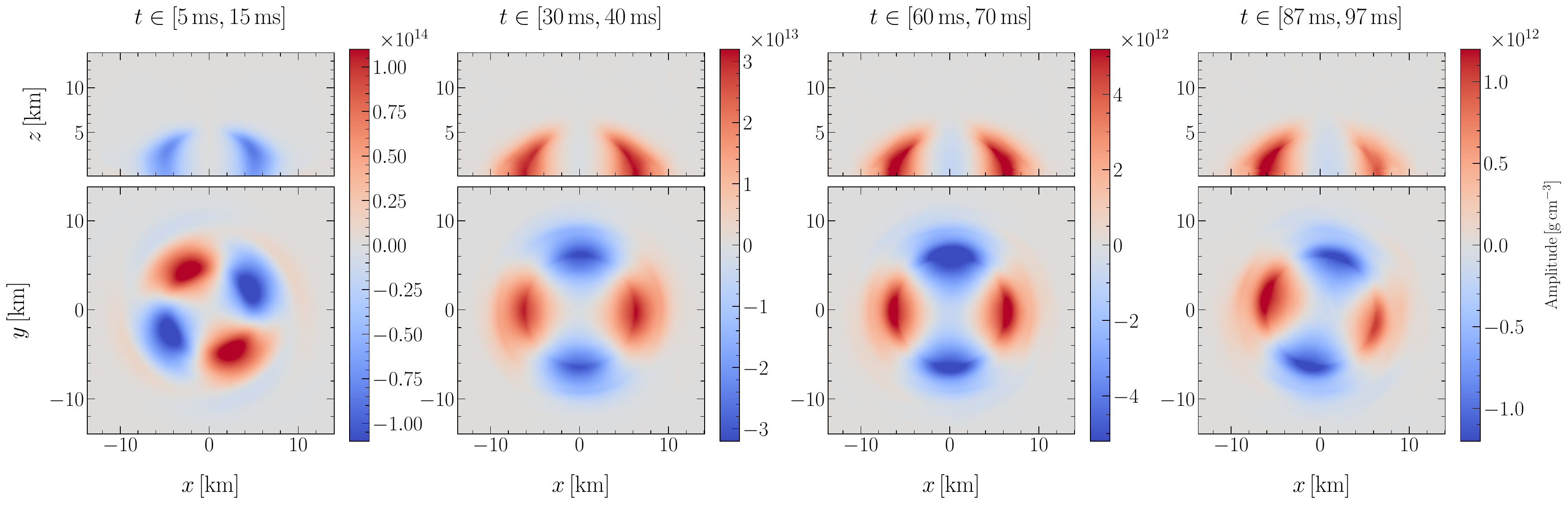}
    \caption{The eigenfunction of the $m=2$ mode in the $x$-$y$ (lower) and $x$-$z$ (upper) planes, respectively, for the \texttt{APR4-135135} model obtained from different post-merger time window (see the top of the figure). Note that the ranges of the colorbars are different for each panel.}
    \label{fig:m2}
\end{figure*}

In Refs.~\cite{DePietri:2018tpx,DePietri:2019mti}, the $ m=2 $ inertial mode is excited following an initial phase of $10$--$20\,\mathrm{ms} $, during which the dominant mode of the massive NS and the primary source of the GW emission is the $ m=2 $ $ f $-mode. This inertial mode excitation lasts for several tens of milliseconds, also emitting GWs. To assess whether inertial modes were excited in our simulations, we perform a Fourier analysis of the $ m=2 $ mass projection, $ C_2 $, and the GW amplitude in the $ l=m=2 $ channel, $ h_{22} $, for the \texttt{APR4-135135} model, as shown in \cref{fig:waveform}. Additionally, in \cref{fig:m2}, we present the eigenfunctions corresponding to the dominant peak observed in \cref{fig:waveform}.

The eigenfunctions validate that the dominant peak appearing in the Fourier spectrum of GWs and matter projections is the $f$-mode. The amplitude of this mode gradually decreases. The oscillations of the NS can be divided into two stages: an early post-merger phase ($ 0 < t - t_{\mathrm{merger}} \lesssim 20\,\mathrm{ms} $), where the GW emission is primarily characterized by the dominant $ m=2 $ $ f $-mode, with subdominant frequencies at the nonlinear coupling frequency of the quadrupolar $f$-mode and quasi-radial oscillations (see Refs.~\cite{Stergioulas:2011gd,Bauswein:2015yca} for details). In the later phase, quasi-radial oscillations dissipate due to hydrodynamic interactions, leaving the $f$-mode as the dominant oscillation in the $m=2$ channel.  

\begin{figure}
    \centering
    \includegraphics[width=\linewidth]{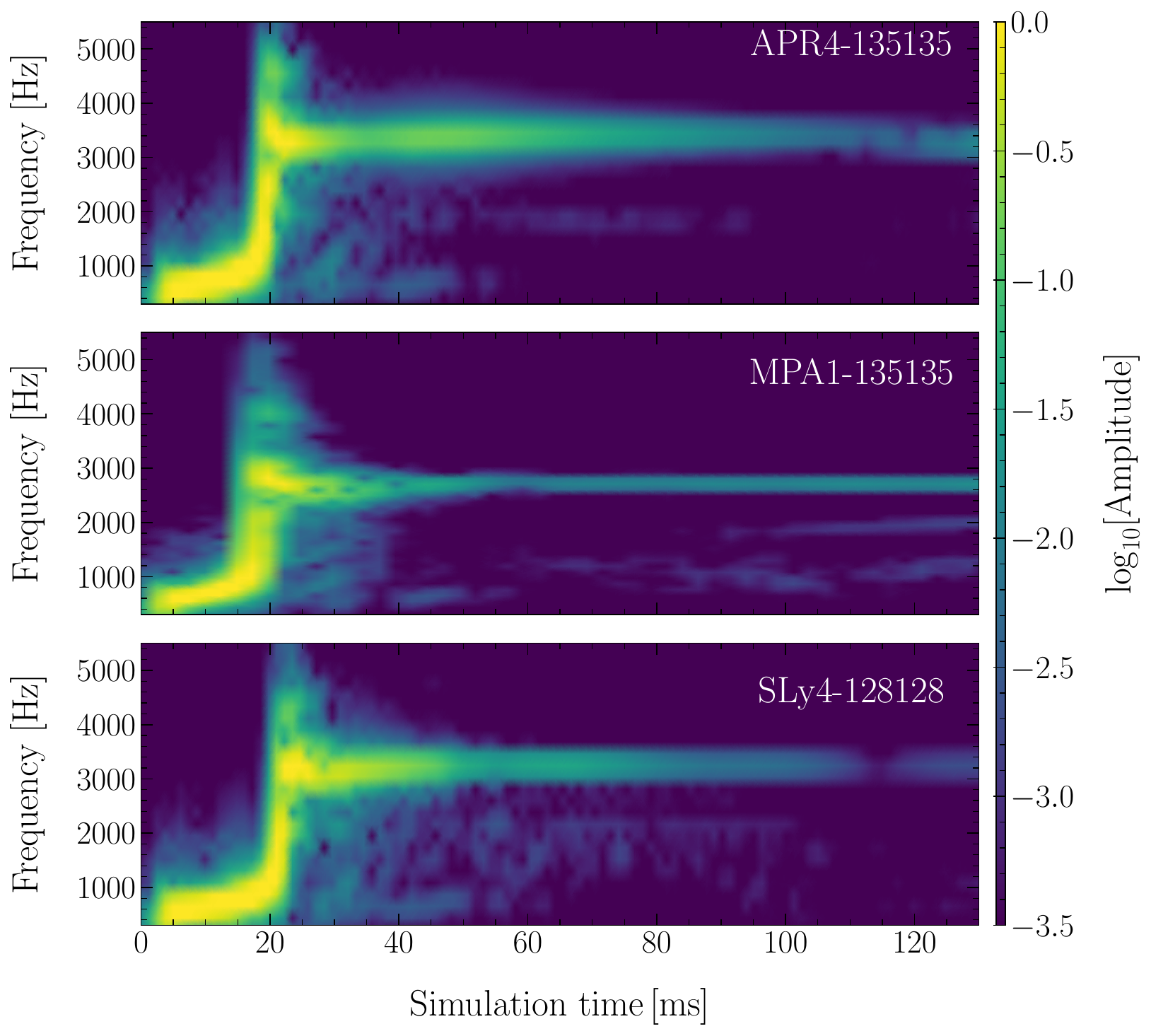}
    \caption{The time-frequency map of $h_{22}$ for three numerical models.}
    \label{fig:spectrogram}
\end{figure}

Comparing the top and bottom panels of Fig.~\ref{fig:waveform}, we observe consistent features shared by the matter projection and GW signals. This alignment supports the accuracy of the quadrupole formula for the GW emission, which has been validated by numerous studies in supernovae~\cite{Reisswig:2010cd,Dimmelmeier:2007ui} and binary NS mergers~\cite{Shibata:2004nv,Shibata:2005gp}.

Fig.~\ref{fig:spectrogram} presents the spectrogram of GWs for the three numerical models. The GW strain exhibits similar characteristics across the models: an early post-merger phase dominated by the $f$-mode with additional side band imprints due to nonlinear couplings, followed by a late post-merger phase where only $f$-mode dominance remains. In the \texttt{MAP1-135135} model, the GW strain shows some low-frequency band features even in the late phase of $t \agt 100$\,ms, but these are 2–-3 orders of magnitude smaller than the $f$-mode component.

Overall, we do not observe strong evidence of inertial mode excitation which was suggested in previous studies in Refs.~\cite{DePietri:2018tpx,DePietri:2019mti,Miravet-Tenes:2024vba}.
If, as suggested by these references, inertial mode excitation is driven by convective instability, our results can be understood by noting that the massive NSs in our simulations do not develop large-scale convective instability regardless of whether we use the Schwarzschild criterion or Criterion I and II, including rotational effects.

\section{Summary and Discussion}
\label{sec:summary}

In this paper, we examine the convective stability and mode characteristics of massive NSs using fully general-relativistic hydrodynamical simulations of BNS mergers and adopting the piecewise APR4, SLy4, and MPA1 EOSs augmented with a $\Gamma$-law thermal component.
These simulations were performed using our latest version of the hydrodynamics code for approximately $120\,\rm ms$ post-merger. Applying the stability criterion for hot, differentially relativistic rotating stars~\citep{Seguin:1975,Bardeen:1970vja}, we find no evidence of large-scale convective instability in the massive NSs, nor any excitation of inertial modes after the quadrupolar $f$-mode damped within $\sim20\,\rm ms$ post-merger. Our results are incompatible with those in Refs.~\cite{DePietri:2018tpx,DePietri:2019mti,Miravet-Tenes:2024vba}, in which the Schwarzschild criterion is violated firstly around $\rho\sim 10^{13\text{--}14}\,\rm g\,cm^{-3}$ and quickly develops to higher density. 

In merger simulations~\cite{DePietri:2018tpx,DePietri:2019mti,Radice:2023zlw} and studies of rotating NSs with nonbarotropic thermal profiles \cite{Camelio:2019rsz}, convective stability is often assessed using only the Schwarzschild criterion. However, given that the merger remnant exhibits rapid rotation and a high degree of differential rotation, the rotational restoring force should also be incorporated into the convective analysis. This necessitates the use of the Solberg-Høiland criterion~\cite{Tassoul2000}, which considers both buoyant and rotational restoring forces. Since GR effects are significant in the NSs, we derive the relativistic Solberg-Høiland criterion for axisymmetric, hot, and differentially rotating fluids, namely Criterion I and Criterion II in \cref{eq:criterion1,eq:criterion2}, following the pioneering work by~\citet{Seguin:1975}, and employ the criterion to the massive NS for the first time. The entropy has positive gradient although the thermal energy exhibits a peak around $\rho \sim 10^{14}\,\rm g\,cm^{-3}$. As a result, the angle-averaged Schwarzschild criterion is satisfied for most regions inside massive NSs. The angle-averaged specific angular momentum increases outward and provides restoring force for perturbed fluid, which makes the massive NSs more stable against convection. The enhancement of convective stability due to rotation has also been observed in black-hole disk~\cite{Sekiguchi:2010ja,Fujibayashi:2020qda} because the specific angular momentum also increases outward.

We also specifically studied the stability on the $x$-$y$ and $x$-$z$ planes. Most regions that have negative entropy gradient/negative gradient of specific angular momentum have been compensated to positive gradient of the specific angular momentum/positive entropy gradient, eliminating the instability. On the $x$-$y$ plane, the region that violates the criterion appears at the shocked region created by the spiral arms, but it quickly changes the location on dynamical timescale, and does not propagate to a large scale. On the $x$-$z$ plane, the density of regions that violates the stability criterion is generally smaller than $\sim 10^{12}\,\rm g\,cm^{-3}$ for our numerical models, have little impact on the global dynamics of the massive NSs.

The inertial modes may be excited by convective instability~\cite{DePietri:2018tpx,DePietri:2019mti}, and have a mixed character of gravity-inertial waves. We found that the inertial modes should be dominated by rotation effects in the high density part of the massive NSs. In the disk region with radii $\varpi \gtrsim 15\,\rm km$, the BV frequency and the epicyclic frequency are comparable, and the inertial modes may have a mixed feature of inertial and gravity waves. Using the mass projection in azimuthal quantum number, we do not observe large-amplitude inertial-mode oscillations. Correspondingly, $l=m=2$ GWs are always dominated by the $f$-mode oscillation. The reason is simply that the massive NSs in our simulations do not develop large-scale convective instability according to the correlation of the convective instability and excitation of inertial modes proposed by Refs.~\cite{DePietri:2018tpx,DePietri:2019mti}.

In addition, we observe an $m=1$ one-armed instability in the post-merger stage for all three numerical models, consistent with previous BNS simulations~\cite{East:2015vix,East:2016zvv,Nedora:2019jhl,Radice:2016gym,Lehner:2016wjg}. We employ the symmetry-preserving technique for the first time in numerical-relativity simulations to demonstrate that our code is free from any bugs that could artificially break $\pi$-symmetry.
It's worthwhile to note that, the growth of $m=1$ mode amplitude after merger correlates strongly with the violation of linear momentum conservation. It is thus possible that this instability has a numerical origin. A linear perturbation analysis is awaited to investigate whether the one-armed spiral instability in the merger remnants is a physical reality. A dedicated numerical technique to ensure linear momentum conservation in simulations is also needed to prevent possible artifacts~\cite{Jiang:2012gy,White:2023wxh}.

\begin{table*}
\centering
\begin{threeparttable}
\caption{The main differences of our work compared to \citet{DePietri:2018tpx,DePietri:2019mti} and \citet{Radice:2023zlw}. Here PWP denotes Piecewise Polytropes; PPM denotes the Piecewise Parabolic Method;
HLLE denotes the Harten-Lax-van Leer-Einfeldt approximate Riemann solver;
WENO5 denotes the fifth-order Weighted Essentially Non-Oscillatory reconstruction;
MP5 denotes the fifth-order Monotonicity-Preserving reconstruction;
DD2 denotes a specific density-dependent relativistic mean-field EOS.}
\begin{tabular}{lccccccc}
\toprule
Literature &
EOS &
\makecell{Evolution\\ code} &
\makecell{Reconstruction \\ at cell interface} &
\makecell{convection \\ criteria\tnote{\textbf{c}}} &
\makecell{convective \\ instability} &
\makecell{inertial \\ modes} \\
\midrule
our work & \makecell{Hybrid EOS \\(4-segments PWP~\cite{Read:2008iy}, \\$\Gamma_{\rm th} =1.8$)}  & \texttt{SACRA-MPI}~\cite{Kiuchi:2017pte,Yamamoto:2008js} & \makecell{HLLC\\$+$PPM}  & \makecell{relativistic\\ Solberg-H{\o}iland}~\cite{Seguin:1975}    &  No & No  \\
\\
Refs.~\cite{DePietri:2019mti,DePietri:2018tpx} & \makecell{Hybrid EOS$\tnote{\textbf{a}}$ \\(7-segments PWP~\cite{Read:2008iy}, \\ $\Gamma_{\rm th} =1.8$)}    & \makecell{\texttt{Einstein}\\
\texttt{Toolkit}~\cite{Loffler:2011ay}\\with $\pi$-symmetry} & \makecell{HLLE\\$+$WENO5} & Schwarzschild~\cite{Thorne:1966ApJ} & Yes & Yes  \\
\\
Ref.~\cite{Radice:2023zlw}   & \makecell{Tabulated EOS\tnote{\textbf{b}} \\ DD2~\cite{Typel:2009sy,Hempel:2009mc} }  & \texttt{THC\_M1}~\cite{Radice:2012cu,Radice:2021jtw} & \makecell{Finite difference\\$+$MP5} & Ledoux~\cite{Ledoux:1947ApJ}  & No   & No \\
\bottomrule
\end{tabular}
\begin{tablenotes}
\footnotesize
\item[a] The numerical simulations are carried out for equal-mass binaries described by the APR4, H4, MS1, and SLy EOSs, each with a component baryonic mass of $M_0 = 1.4\,M_{\odot}$. The finest grid resolution is $\Delta x_{\rm finest} = 136\,\rm m$ in the \texttt{SLy-128128} model. 
\item[b] The neutrino-radiation hydrodynamic simulations are carried out for an equal-mass binary with the component mass $M=1.35\,M_{\odot}$. 
\item[c] The relativistic Solberg–H{\o}iland criterion used in this work accounts for both differential rotation and thermal gradients. In contrast, the Schwarzschild criterion adopted in Refs.~\cite{DePietri:2018tpx,DePietri:2019mti} includes only thermal gradients. Both literature neglect compositional gradients because the simulations employ hybrid EOSs. Ref.~\cite{Radice:2023zlw} uses the Ledoux criterion, which incorporates both thermal and compositional gradients but ignores rotational effects.
\end{tablenotes}
\label{tab:compare}
\end{threeparttable}
\end{table*}

In the present work, we adopt a $\Gamma$-law thermal component and neglect the effects of microphysics and magnetic fields for simplicity. However, these assumptions may influence our conclusions. In reality, $\Gamma_{\rm th}$ may depend strongly on density and temperature~\cite{Keller:2020qhx,Keller:2022crb,Raithel:2019gws,Raithel:2023zml,Constantinou:2015mna}, which could significantly impact the thermal buoyancy within the massive NSs. For instance, if $\Gamma_{\rm th}$ is substantially larger around the `hot ring' with temperature peaks compared to the outer regions, as shown in \cref{fig:contour_reo,fig:omega_T_S}, it might lead to a negative entropy gradient. However, we are uncertain about stabilizing effects of rotation under such conditions. Including microphysics would introduce buoyancy effects due to compositional gradients, which can be incorporated into our criteria directly. \citet{Radice:2023zlw} examined the convective instability of the massive NSs in general-relativistic neutrino-radiation hydrodynamics simulations of BNS mergers. Using the Ledoux criterion, which accounts for both thermal and compositional gradients, they concluded that the star is stable against convection. 

{We have discussed how our physical results differ from those of Refs.~\cite{DePietri:2018tpx,DePietri:2019mti,Radice:2023zlw}. As emphasized in the Introduction, the present work serves as an independent study rather than a one-to-one comparison. Nevertheless, it is useful for the reader to have a concise overview of the differences among these works. For this reason, we include in Tab.~\ref{tab:compare} a summary of the key distinctions, such as the adopted EOS, evolution codes, reconstruction of hydro-variables at the cell interface, convection criteria, and main results. }

The influence of magnetic fields on the convective stability is not clear. One established effect is that effective viscosity from MHD turbulence rapidly eliminates differential rotation in the massive NSs on the viscous timescale. Refs.~\cite{Shibata:2017xht,Fujibayashi:2017puw} modeled this effective viscosity using the viscous $\alpha$-parameter, showing that differential rotation in the massive NSs is significantly reduced within $\lesssim 5\,\mathrm{ms}$ if $\alpha \sim \mathcal{O}(10^{-2})$. As a result, the core of the massive NS transitions to nearly uniform rotation, which has a stabilizing effect against convection due to the outwardly increasing specific angular momentum in uniform rotation. Future research incorporating more realistic thermal behavior, microphysics, and magnetic fields is essential to give a more definitive conclusion.

\section*{Acknowledgements}

We thank the members of the Computational Relativistic Astrophysics group in AEI for helpful discussions. {We also thank José Antonio Font, Sebastiano Bernuzzi, and Milton Ruiz for useful discussions during the conference New Frontiers NR 2025.} Yong Gao thanks Alexis Reboul-Salze and Shichao Wu for useful discussions. Numerical computations were performed on the clusters Sakura, Cobra, and Raven at the Max Planck Computing and Data Facility. This work was in part supported by Grant-in-Aid for Scientific Research (grant Nos. 20H00158, 23H04900, 23H01172 and 23K25869) of Japanese MEXT/JSPS.


\bibliography{refs}

\appendix

\section{Details on Symmetry-Preserving Techniques}
\label{sec:appendixA}

This appendix describes a symmetry-preserving technique with which the $\pi$-symmetry is preserved to machine precision in the symmetry-preserving test.

\subsection{Prolongation}

\begin{figure*}
    \centering
    \includegraphics[width=\linewidth]{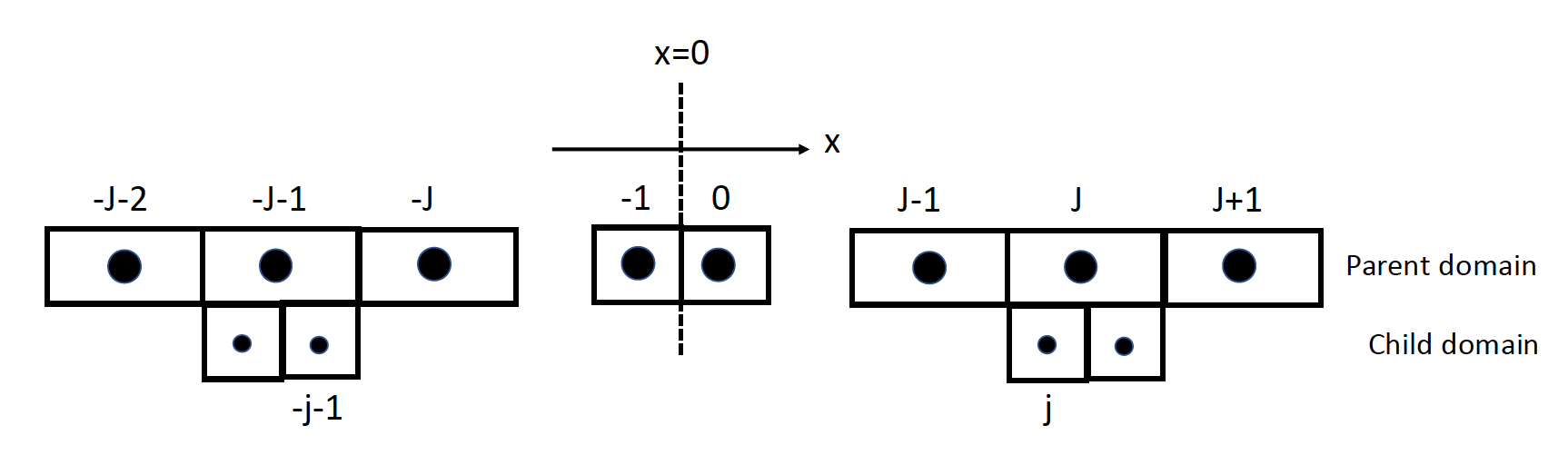}
    \caption{Schematic picture of prolongation in a cell-center nested grid. $J$ and $j$ denote the grid indices in the $x$-direction in a
    parent domain and child domain, respectively.}
    \label{fig:prolongation}
\end{figure*}

We need a prolongation with the mesh refinement technique, i.e., the interpolation from a coarser nested grid domain to a finer nested grid domain. For example, for simplicity, we consider the second-order Lagrange interpolation in the staggered grid in one dimension. Generalization to a higher-order Lagrange prolongation or the prolongation with a limiter function such as WENO5 is straightforward. Also, for simplicity, we assume the coordinate origins for the nested domains coincide, which is not necessarily the case in simulations.
\cref{fig:prolongation} plots a schematic picture for the prolongation at $j$ and $-j-1$ in a child (finer) nested grid domain. $J$ and $-J-1$ denote the grid index in a parent (coarser) nested grid domain. Normally, a scalar quantity $q$ at $j$ and $-j-1$ in the child domain is interpolated by
\begin{align}
    \label{eq:A1}
    & q_j=\frac{5}{32} Q_{J-1}+\frac{15}{16} Q_J-\frac{3}{32} Q_{J+1} \,,\\
    \label{eq:A2}
    & q_{-j-1}=-\frac{3}{32} Q_{-J-2}+\frac{15}{16} Q_{-J-1}+\frac{5}{32} Q_{-J}\,,
\end{align}
in the code with the “do” loop in Fortran or the “for” loop in C in terms of $j$ where $Q$ is the data in the parent domain. However, this way of the prolongation does not 
guarantee that $q_j$ and $q_{-j-1}$ agree to machine precision even if $Q$ satisfies a symmetric condition, i.e., $Q_{J-1}=$ $Q_{-J}, Q_J=Q_{-J-1}$, and $Q_{J+1}=Q_{-J-2}$ because the adding order is different in \cref{eq:A1,eq:A2}. In our code, we implement the prolongation for $q_{-j-1}$ by
\begin{equation}
q_{-j-1}=\frac{5}{32} Q_{-J}+\frac{15}{16} Q_{-J-1}-\frac{3}{32} Q_{-J-2}
\end{equation}
to make the adding order the same as in \cref{eq:A1}. With this prescription, $q_j=q_{-j-1}$ for the scalar quantity $q$ is guaranteed to machine precision in the symmetry-preserving test if $Q$ satisfies the symmetric condition. It also holds for the vector and tensor quantities except for
the sign.

\subsection{Riemann solver}

We implement a similar technique to our Riemann solver. For example, the HLLC Riemann solver requires to calculate the Harten-Lax-Lee state \cite{Kiuchi:2022ubj,Mignone:2005ft} :
\begin{equation}
Q_{j+\frac{1}{2}}^{\mathrm{HLL}}=\left\{\frac{\lambda_R Q^R-\lambda_L Q^L+F\left(Q^L\right)-F\left(Q^R\right)}{\lambda_R-\lambda_L}\right\}_{j+\frac{1}{2}}\,,
\end{equation}
where $\lambda_{L / R}, Q^{L / R}$, and $F(Q)$ are the characteristic speed, the conserved quantity, and physical flux in the left/right state at the cell interface of $x_{j+\frac{1}{2}}$ (see Ref.~\cite{Kiuchi:2022ubj} in detail), respectively. In our implementation, we change the order of the calculation for $j<0$ by
\begin{equation}
Q_{j-\frac{1}{2}}^{\mathrm{HLL}}=\left\{\frac{-\lambda_L Q^L+\lambda_R Q^R-F\left(Q^R\right)+F\left(Q^L\right)}{-\lambda_L+\lambda_R}\right\}_{j-\frac{1}{2}} \,.
\end{equation}
With it, it is guaranteed that $Q_{j+\frac{1}{2}}^{\mathrm{HLL}}=Q_{-j-\frac{3}{2}}^{\mathrm{HLL}}$ for the scalar quantity $Q$ to machine precision in the symmetry-preserving test. It is also guaranteed for the vector quantity except for the sign. Thus, the numerical flux obtained by the Riemann problem also satisfies this property.

\section{The linear momentum violation in mesh refinement}

\label{sec:appendixB}
The extent to which the linear momentum is conserved in a simulation is crucial. In this Appendix, we demonstrate that any numerical relativity
simulations with a mesh-refinement technique, such as the nested grid, cannot avoid the violation of linear momentum conservation. Let us consider the discretized form of the equation of
motion of the x-component of the momentum:
\begin{widetext}
\begin{align}
    \label{eq:linear_diff}
 \partial_t\left(S_x\right)_{j, k, l}+\frac{\left(\sqrt{\gamma} \tilde{F}_x^x\right)_{j+\frac{1}{2}, k, l}-\left(\sqrt{\gamma} \tilde{F}_x^x\right)_{j-\frac{1}{2}, k, l}}{\Delta x}+\frac{\left(\sqrt{\gamma} \tilde{F}_x^y\right)_{j, k+\frac{1}{2}, l}-\left(\sqrt{\gamma} \tilde{F}_x^y\right)_{j, k-\frac{1}{2}, l}}{\Delta y}+\frac{\left(\sqrt{\gamma} \tilde{F}_x^z\right)_{j, k, l+\frac{1}{2}}-\left(\sqrt{\gamma} \tilde{F}_x^z\right)_{j, k, l-\frac{1}{2}}}{\Delta z} 
 =\left(\bar{M}_x\right)_{j, k, l}\,,
\end{align}
\end{widetext}
where $S_x=\sqrt{\gamma} \rho W h u_x$ and $\tilde{F}_x^i\,(i=x,\ y,\ z)$ is a numerical flux. $\bar{M}_x$ is a source term (see Ref.~\cite{Kiuchi:2022ubj} for the detailed finite volume method, and the concrete expressions for the numerical flux and source terms). The indices $j,\ k$, and $l$ denote the grid point in the $x$-, $y$-, and $z$-directions, respectively, e.g., $x_j=\left(j+\frac{1}{2}\right) \Delta x$ with $j=-N,\  \cdots,\  N-1$.

As a representative term in the source $\bar{M}_x$, we pick up $-S_0 \partial_x \alpha$ where $S_0=\sqrt{\gamma} \rho W\left(h W-\frac{p}{\rho W}\right)$. The following discussion can be applied to the other terms in $\bar{M}_x$. First, we consider a single numerical domain, i.e., no mesh refinement. We integrate \cref{eq:linear_diff} by
\begin{widetext}
\begin{align}
    \label{eq:linear_int}
& \sum_{j, k, l=-N}^{N-1} \partial_t\left(S_x\right)_j \Delta x \Delta y \Delta z+\sum_{j, k, l=-N}^{N-1}\left\{\left(\sqrt{\gamma} \tilde{F}_x^x\right)_{j+\frac{1}{2}, k, l}-\left(\sqrt{\gamma} \tilde{F}_x^x\right)_{j-\frac{1}{2}, k, l}\right\} \Delta y \Delta z \nonumber\\
& +\sum_{j, k, l=-N}^{N-1}\left\{\left(\sqrt{\gamma} \tilde{F}_x^y\right)_{j, k+\frac{1}{2}, l}-\left(\sqrt{\gamma} \tilde{F}_x^y\right)_{j, k-\frac{1}{2}, l}\right\} \Delta x \Delta z+\sum_{j, k, l=-N}^{N-1}\left\{\left(\sqrt{\gamma} \tilde{F}_x^z\right)_{j, k, l+\frac{1}{2}}-\left(\sqrt{\gamma} \tilde{F}_x^z\right)_{j, k, l-\frac{1}{2}}\right\} \Delta x \Delta y \nonumber\\
& =-\sum_{j, k, l=-N}^{N-1}\left(S_0\right)_{j, k, l} \frac{\alpha_{j+1, k, l}-\alpha_{j-1, k, l}}{2} \Delta y \Delta z+\text { (the other terms) }\,.
\end{align}
\end{widetext}
where we adopt the second-order finite difference expression for $\partial_x \alpha$, but the generalization to higher-order finite difference is straightforward. The second, third, and fourth terms on the left-hand side vanish where we assume no flux at the boundary, e.g., $\left(\tilde{F}_x^x\right)_{-N-1 / 2, k, l}=0=$ $\left(\tilde{F}_x^x\right)_{N-1 / 2, k, l}$. The sum in the right-hand side is decomposed into
\begin{widetext}
\begin{align}
& (\text { R.H.S })=-\sum_{l=-N}^{N-1}\left(\sum_{k=-N}^{-1}+\sum_{k=0}^{N-1}\right)\left(\sum_{j=-N}^{-1}+\sum_{j=0}^{N-1}\right)\left(S_0\right)_{j, k, l} \frac{\alpha_{j+1, k, l}-\alpha_{j-1, k, l}}{2} \Delta y \Delta z+(\text { the other terms) } \nonumber\\
& =-\sum_{l=-N}^{N-1}\left[\sum_{k^{\prime}=0}^{N-1} \sum_{j^{\prime}=0}^{N-1}\left(S_0\right)_{-j^{\prime}-1,-k^{\prime}-1, l} \frac{\alpha_{-j^{\prime},-k^{\prime}-1, l}-\alpha_{-j^{\prime}-2,-k^{\prime}-1, l}}{2}+\sum_{k=0}^{N-1} \sum_{j^{\prime}=0}^{N-1}\left(S_0\right)_{-j^{\prime}-1, k, l} \frac{\alpha_{-j^{\prime}, k, l}-\alpha_{-j^{\prime}-2, k, l}}{2}\right. \nonumber\\
& \left.+\sum_{k^{\prime}=0}^{N-1} \sum_{j=0}^{N-1}\left(S_0\right)_{j,-k^{\prime}-1, l} \frac{\alpha_{j+1,-k^{\prime}-1, l}-\alpha_{j-1,-k^{\prime}-1, l}}{2}+\sum_{k=0}^{N-1} \sum_{j=0}^{N-1}\left(S_0\right)_{j, k, l} \frac{\alpha_{j+1, k, l}-\alpha_{j-1, k, l}}{2}\right] \Delta y \Delta z+\text { (the other terms) } \nonumber\\
& =-\sum_{l=-N}^{N-1}\left[\sum_{k^{\prime}=0}^{N-1} \sum_{j^{\prime}=0}^{N-1}\left(S_0\right)_{j^{\prime}, k^{\prime}, l} \frac{\alpha_{j^{\prime}-1, k^{\prime}, l}-\alpha_{j^{\prime}+1, k^{\prime}, l}}{2}+\sum_{k=0}^{N-1} \sum_{j^{\prime}=0}^{N-1}\left(S_0\right)_{j^{\prime},-k-1, l} \frac{\alpha_{j^{\prime}-1,-k-1, l}-\alpha_{j^{\prime}+1,-k-1, l}}{2}\right. \nonumber\\
& \left.+\sum_{k^{\prime}=0}^{N-1} \sum_{j=0}^{N-1}\left(S_0\right)_{j,-k^{\prime}-1, l} \frac{\alpha_{j+1,-k^{\prime}-1, l}-\alpha_{j-1,-k^{\prime}-1, l}}{2}+\sum_{k=0}^{N-1} \sum_{j=0}^{N-1}\left(S_0\right)_{j, k, l} \frac{\alpha_{j+1, k, l}-\alpha_{j-1, k, l}}{2}\right] \Delta y \Delta z+(\text { the other terms) } \nonumber\\\nonumber\\
& =0\,,
\end{align}
\end{widetext}
where we replace the index from $j$ (resp. $k$ ) to $j^{\prime}=-j-1$ (resp. $k^{\prime}=-k-1$ ) in the second equality and we use a $\pi$-symmetry condition $Q_{j^{\prime}, k^{\prime}, l}=Q_{-j^{\prime}-1,-k^{\prime}-1, l}$ where $Q=S_0$ or $\alpha$ in the third equality. The important point is that the $\pi$-symmetry condition is realized to machine precision in the symmetry-preserving test described above. As a consequence, the volume integration of $S_x$ is preserved, i.e., the linear momentum conservation. This proof also holds in a mesh refinement case. Note that we need the reflux prescription for the second, third, and fourth terms on the left-hand side of \cref{eq:linear_int} to be canceled out.

However, in general~\footnote{For example, with an optimization compile option without \texttt{-fp-model strict}.}, this is not guaranteed because the $\pi$-symmetry condition is not realized numerically due to noise, e.g., the round-off error, even if we employ an ID which possesses the $\pi$-symmetry and the symmetry-preserving technique described in Appendix A. In particular, when the source term $\bar{M}_x$ is evaluated, the finite difference error abruptly changes at the refinement boundary due to the $2: 1$ refinement algorithm. Therefore, the noise is significantly enhanced near the refinement boundary. As a result, the violation of the $\pi$-symmetry condition is significant there and leads to the violation of linear momentum conservation.

\section{Resolution study for the \texttt{APR4-135135} model}
\label{sec:AppendixC}

In this Appendix, we present the resolution study of the $m=1$ one-armed instability, criteria of the convective stability, and the GW spectrum for the \texttt{APR4-135135} model. 

\begin{figure}
    \centering
    \includegraphics[width=\linewidth]{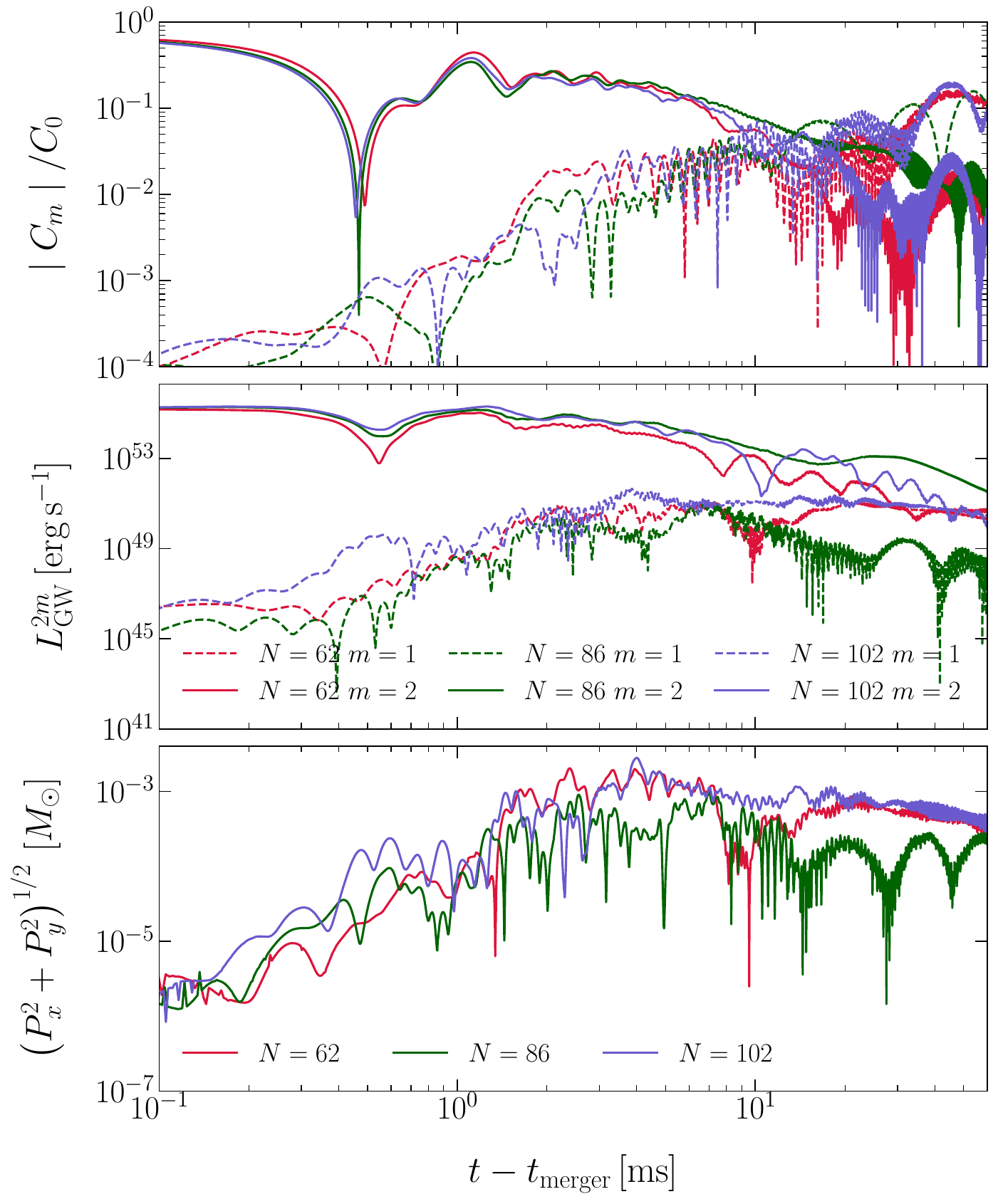}
    \caption{The post-merger time evolution of quantities for both $m = 1$ and $m = 2$ modes across the three different resolutions for the model \texttt{APR4-135135}: the normalized density projection, ${|C_{m}|}/{C_{0}}$ (top), the GW luminosity, $L_{\rm GW}^{2m}$ (middle), and the linear momentum violation, $\sqrt{P_x^2 + P_y^2}$ (bottom). The top and middle panels share the same labels. Note that the post-merger time for the middle panel is the retarded value at $r_{0}= 480\, M_{\odot}$.}
    \label{fig:resolution_projection}
\end{figure}

In \cref{fig:resolution_projection}, we show the density projection $C_{m}$, the GW luminosity $L_{\rm GW}^{2m}$, and the linear momentum violation for three different grid resolutions: $N=62, 86, \text{and}\, 102$. The amplitude of the $m=1$ mode and the linear momentum violation exhibit exponential growth after the merger, independent of the grid resolution. However, compared to the canonical resolution of $N=86$, the $m=1$ mode amplitude reaches a higher saturation level for the cases with $N=62$ and $N=102$.

\begin{figure}
    \centering
    \includegraphics[width=\linewidth]{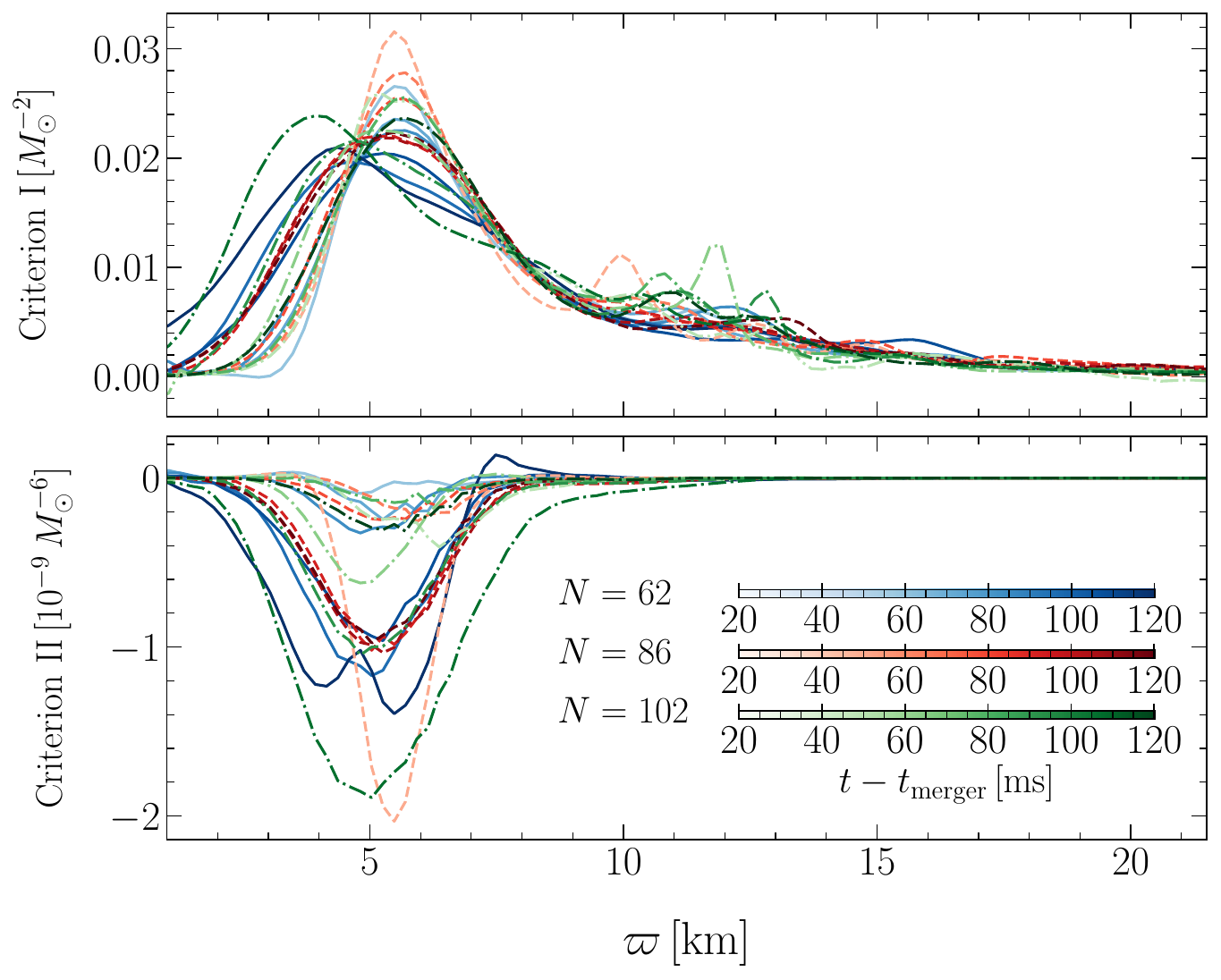}
    \caption{The post-merger time evolution of the Criterion I (upper) and Criterion II (lower) of the convective stability for the three different resolutions with $N=62$, $N=86$, and $N=102$.}
    \label{fig:resolution_critI_II}
\end{figure}

\begin{figure}
    \centering
    \includegraphics[width=\linewidth]{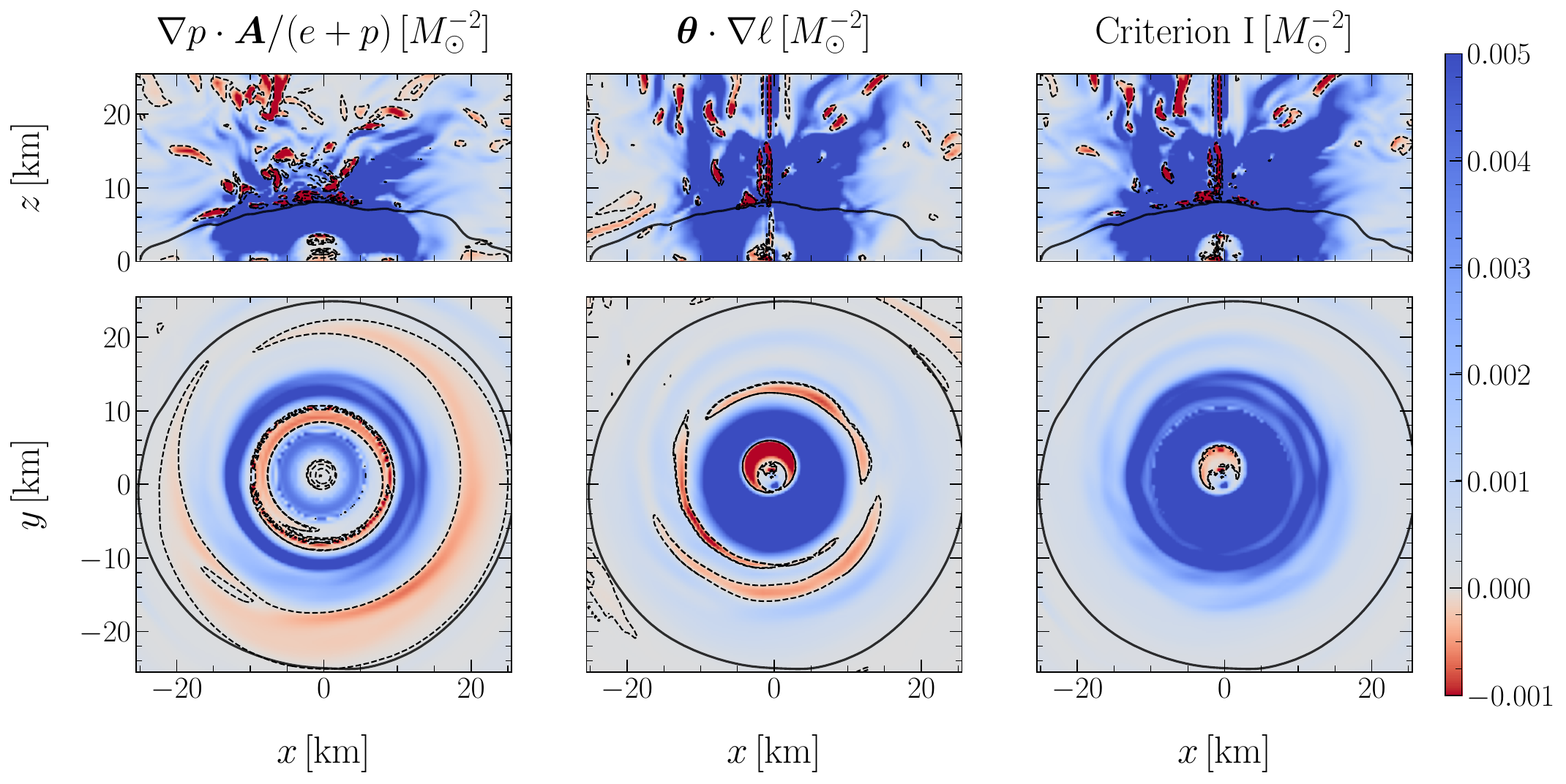}
    \caption{Same as \cref{fig:critI_snapshot} but for the case with grid number $N=62$.}
    \label{fig:resolution_N62}
\end{figure}

\begin{figure}
    \centering
    \includegraphics[width=\linewidth]{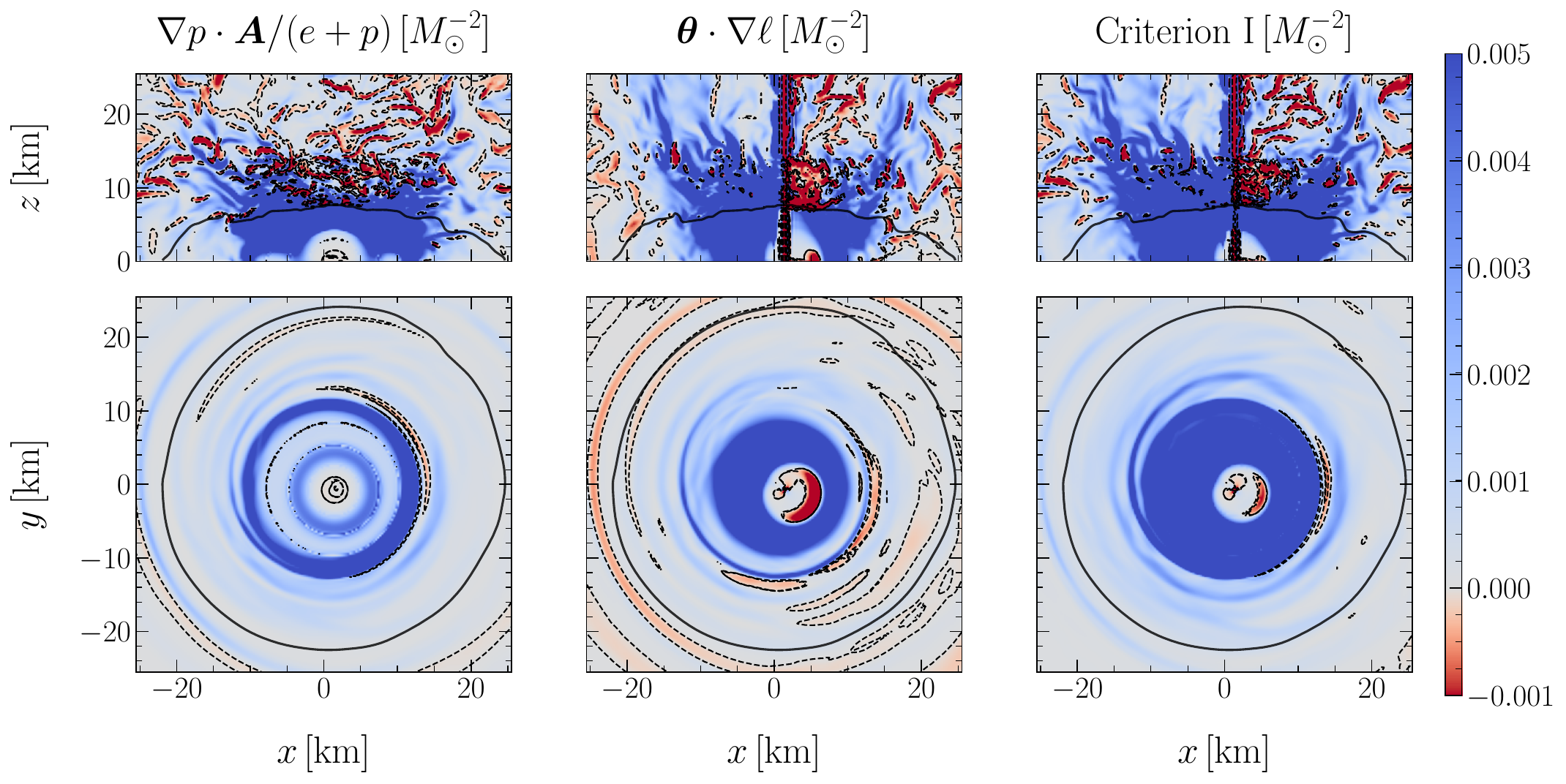}
    \caption{Same as \cref{fig:critI_snapshot} but for the case with grid number $N=102$.}
    \label{fig:resolution_N102}
\end{figure}

The angle-averaged Criterion I and Criterion II on the equatorial plane for three resolutions are presented in \cref{fig:resolution_critI_II}. Both criteria are largely maintained across most regions at different post-merger times. \Cref{fig:resolution_N62,fig:resolution_N102} further illustrate snapshots of Criterion I on the $x$-$y$ and $x$-$z$ planes for grid resolutions $N=62$ and $N=102$. The results are consistent with the $N=86$ case, except for a larger violation of the criterion at the center of the massive NS. This discrepancy arises from the larger amplitude of the $m=1$ mode at that moment for $N=62$ and $N=102$, leading to a non-axisymmetric structure in both the density and rotational profiles. 

\begin{figure}
    \centering
    \includegraphics[width=\linewidth]{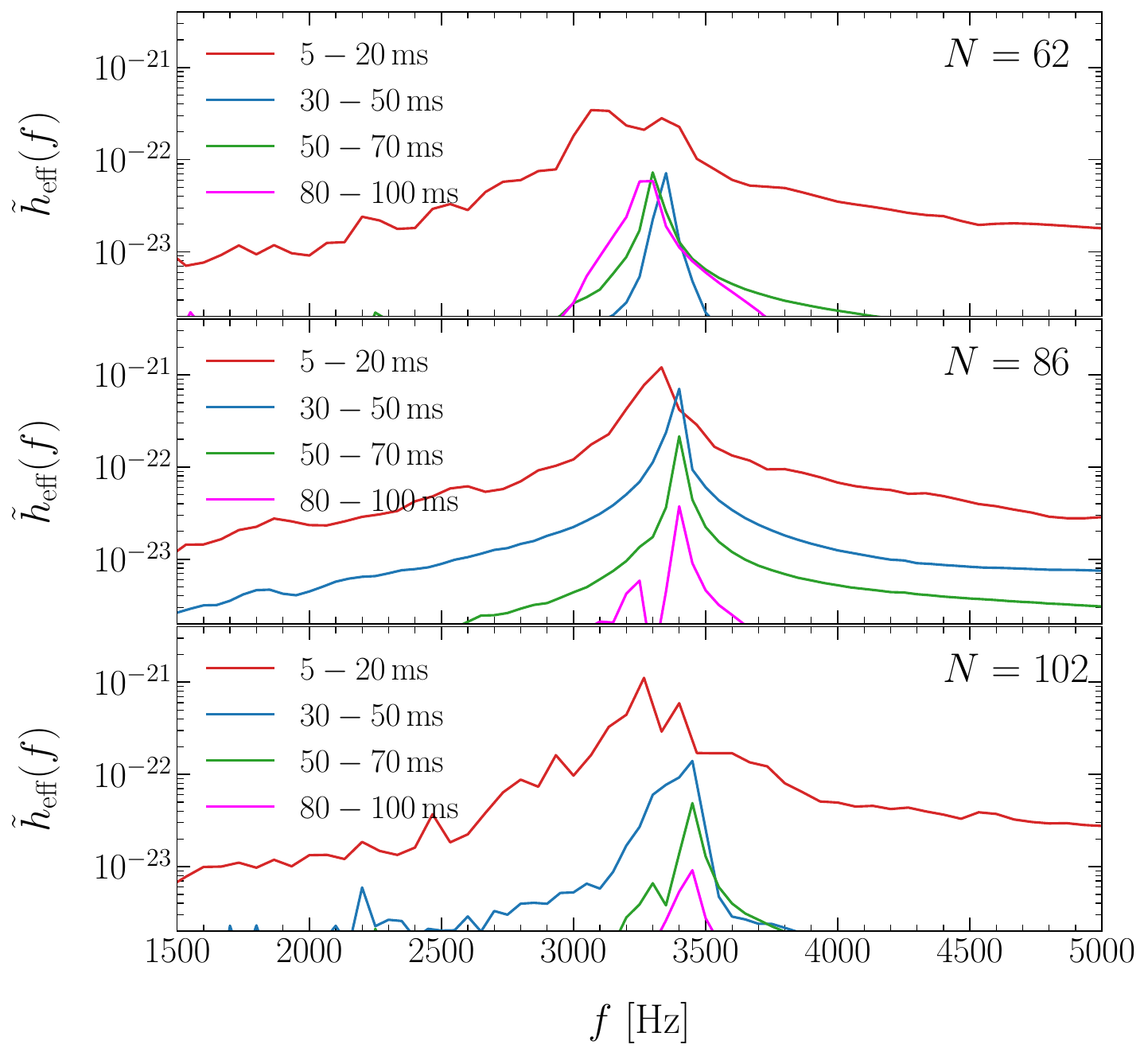}
    \caption{The ASD of the effective amplitude $\tilde{h}_{\rm eff}(f)$ for different resolutions with $N=62$, $N=86$, and $N=102$ at different post-merger time window. The luminosity distance is taken as $50\,\rm Mpc$.}
    \label{fig:resolution_GW}
\end{figure}

In \cref{fig:resolution_GW}, we show the ASD of GWs in the $m=2$ channel for the three resolutions. The GWs are consistently dominated by the $f$-mode, without excitation of other modes. The frequency of the $f$-mode gradually increases over time due to the contraction of the massive NS. While it can be observed that the peak frequency migrates to a lower value at post-merger time $t-t_{\rm merger} \sim 80 \,\rm ms$ for $N=62$, the low-resolution results might not be very reliable in the late stage of the evolution.

\clearpage

\end{document}